\documentstyle[pre,aps,epsf,psfig]{revtex}
\newcommand \be{\begin{equation}}
\newcommand \ee{\end{equation}}
\newcommand \beq{\begin{eqnarray}}
\newcommand \eeq{\end{eqnarray}}

\begin{document}
\draft
\title{
Equation of states for classical Coulomb systems.\\ 
The use of Hubbard-Schofield approach.
}

\author{J. Ortner}
\address{Institut f\"ur Physik,
Humboldt Universit\"at Berlin\\
Invalidenstr. 110, 10115 Berlin, Germany
}
\date{to be published in Phys. Rev. E}
\maketitle
\begin{abstract}
An effective method based on Hubbard-Schofield approach [Phys. Lett. A {\bf 40}, 245 (1972)] is developed to calculate the free energy of classical Coulomb systems. This method significantly simplifies the derivation of the cluster expansion.  A diagrammatic representation of the cluster integrals is proposed. Simple rules providing the leading order in density $n$ of each diagrammatic contribution are found. We calculate the $n^3$ contribution and recover the results at the order $n^{5/2}$ obtained by the traditional method of resummation of diverging Mayer bonds. 
\end{abstract}

\pacs{PACS numbers:05.20.Gg, 05.70.Ce, 52.25.Kn}

\section{Introduction}
This paper is adressed to the study of the virial expansion of the Helmholtz free energy (hereafter the free energy) for a classical Coulomb system. We consider a multicomponent system of point-like ions embedded in a neutralizing background. Because of the long-range Coulombic nature of the interaction potential between two charges the corresponding virial expansion Mayer graphs \cite{M40} diverge. In the traditional method these long-range divergencies are removed via the chain resummations first introduced by Mayer \cite{M50} and Salpeter \cite{S58} and further developed by the works of Meeron \cite{Meeron58} and Abe \cite{A59}. To avoid complicated calculations related to the chain resummation we present in this paper an alternative method of calculation of classical Coulomb systems thermodynamic functions. 

We show in this paper that the earlier results can be obtained much easier using the method of collective variables and integral transformations. Moreover,  this method makes it possible to obtain the free energy of a classical plasma system in a systematic manner up to an arbitrary order (at least in principal). In addition, this method is capable to describe  not only the low density limiting thermodynamic behavior of a Coulomb system but also the region of critical density of Coulomb fluids and that of a strongly coupled plasma. However, in this paper we will restrict our considerations to the low density region of a classical ion mixture. 

The method of collective variables is a powerfool tool of investigation of both classical and quantum Coulomb systems. There were two basic lines in this method mainly developed already in the fifties. The first line starts with the initial plasma Hamiltonian being the sum of charges kinetic energy and the interaction potential between them and converts it into the collective variables Hamiltonian using the canonical transformations   \cite{BP51}.

The second line \cite{Z54,J58} starts from the configurational integral of individual particles. Certain transformations lead then to the configurational integral expressed through collective variables.

Using the first approach the physical processes can be interpreted in terms of the  collective variables. Thus the plasma oscillations are represented as oscillations of the density fluctuations Fourier components. Following the second general line it is possible to obtain an expansion of the cluster integral in a systematic manner.

Using the formal representation of the configurational integral in terms of the collective variables Zubarev \cite{Z54} and Juchnovskij \cite{J58} gave a cluster expansion of the free energy. The transformation from the individual (i.e., the position of the particles) to the collective coordinates (i.e., the Fourier transforms of the density fluctuations) is performed via the corresponding Jacobian. Within the Random Phase Approximation for the Jacobian one arrives at the Debye-H\"uckel approximation for the free energy \cite{Z54}, and a systematic expansion of the Jacobian leads to the cluster expansion of the free energy \cite{J58,JG80}. However, this method is still cumbersome.

By use of integral transformations the method of collective variables have become technically more feasible. Here, instead of a Jacobian transformation an identity is used which expresses the Coulomb interaction in terms of external interactions \cite{Stratonovich57,Hubbard58,K59}.  As for Coulomb systems this exact version of the mean field idea of Debye-H\"uckel leads to the sine-Gordon (SG) representation of the configurational integral \cite{K59,S60,GK81}. By virtue of a rigorous mathematical proof it was shown that the SG transformation produces a cluster expansion of a system with long-range interactions \cite{Brydges}. Further the SG theory of Coulomb gas was used to analyze the metal-insulator transition \cite{Saito,DYL97} and Coulomb criticality \cite{Kh86}. Such a way of analysis of the Coulomb criticality was claimed to be the most promissing \cite{Fisher94}. However, the SG representation in the pure Coulomb version is applicable only for lattice models (with possibly vanishing lattice constant) and for pointlike charges.

A hybrid method combining the advantages of the Zubarev-Juchnovskij approach with that of Stratonovich, Hubbard and Kac was developed in Ref. \cite{HSch72}. Here the particle interactions are divided into long- and short-range interactions.  Thus short-range repulsive interactions as, e.g., hard core repulsions are introduced in a natural way. Using the Stratonovich-Hubbard-Kac transformation the original system with the both types of interactions is mapped onto a reference system with short range interactions only. In contrast to the SG theory the Hubbard-Schofield (HS) approach is capable to perform off-lattice calculations. Further in the HS approach the cumulant expansion is used to map the Hamiltonian of the original system (a nonionic or ionic fluid) onto an effective magnetic-like Hamiltonian.  Since the magnetic system Hamiltonian can be recasted into the Landau-Ginzburg-Wilson form \cite{FL96} such a representation is very convenient for the analysis  of both ordinary fluids \cite{BV98} and charged hard sphere systems criticality \cite{BBB98}. To our knowledge in Ref. \cite{Brilliantov} the HS approach was first applied to a one-component plasma. 

Despite the significant success of the method of collective coordinates in describing Coulomb systems the low density limit of Coulomb systems was mainly studied by traditional methods of statistical physics \cite{F62}. Starting by the charging formula for the free energy an expansion in terms of ordinary Mayer functions can be found. However, in the case of Coulomb systems the Mayer functions diverge. Expanding the Mayer function in powers of the Coulomb potential one can collect the Mayer series into special subseries and perform a partial summation of Mayer bonds. The resummed Mayer bond is representable through the screened Coulomb potential and is integrable. Thus the sum of all ring diagrams gives the Debye-H\"uckel approximation for the free energy. Picking up in all diagrams the chain of Mayer bonds leading to the screened potential and ordering all diagrams by the number of vortices the cluster expansion for the free energy is obtained. The virial expansion up to the order $n^{5/2}$ (n being the number density of ions) was found within this method \cite{A59,Haga,CM59,F62}. However, to perform this procedure a number of very refined diagrammatic transformations are required \cite{F62}.
Various first-principles formalisms based on the chain resummation of Mayer bonds have been used to generalize the virial expansion of classical Coulomb systems to the case of quantum plasmas. First, the method of effective potentials introduced by Morita \cite{Morita} for quantum systems, has been applied by Ebeling to the Coulomb case \cite{Ebeling}.  A renewed interest in the exact calculation of thermodynamic functions for weakly coupled quantum plasmas beyond the Debye-H\"uckel limiting law has been observed recently. 
The virial expansion for the free energy up to the order $n^{5/2}$ in the density was derived using the Feynman-Kac formalism \cite{Alastuey} and the method of Green's functions  \cite{Riemann}, respectively. The latter method involves an additional expansion with respect to the square of charge $e^2$. Confirming and completing the earlier results by Haga \cite{Haga}, Friedman \cite{F62}, DeWitt \cite{DeWitt} and Ebeling \cite{Ebeling} the both approaches arrived at the same result. Therefore the virial equation of state for both classical and quantum Coulomb systems is now generally accepted as known up to order $n^{5/2}$.

In this paper we show that the density expansion of a classical Coulomb system can be obtained much easier using the method of collective coordinates. As Zubarev and Juchnovskij we rewrite the partition function in terms of the collective variables. In contrast to them we will not use the Jacobian transforming the space coordinates of individual particles into the collective variables. Instead, we represent the configurational integral of the Coulomb system through the configurational integral of a noninteracting system in external fields. Further following the line of Hubbard and Schofield we employ the cumulant expansion. This enables us to rewrite the configurational integral in a similar way as the partition function for the magnetic system having an Ising-like Hamiltonian. The coefficients of the effective magnetic-like Hamiltonian are expressable in terms of the ideal gas correlation functions and are perfectly known. Performing a perturbation expansion in terms of the anharmonic contributions to the effective Hamiltonian one directly obtains the cluster expansion of the configurational integral and of the free energy of a classical plasma without bonds resummation (Section \ref{S1}). Instead the bare Coulomb interaction is automatically screened in the HS approach. Ultimately, the integrals that determine the virial coefficients in the HS method are the same which appear in the more traditional Abe-Meeron-Friedmann approach. In this sense the HS method may be considered as another evidence of the fact that all virial coefficient are expressable in terms of Mayer-like graphs built with screened bonds. Arriving at the virial equation of state it will be shown that the method of collective variables in the Hubbard-Schofield representation is capable to describe not only the critical behavior but also the low-density limit of Coulomb systems. 

Additionally, there are some basic ideas in the literature on how to generalize the method of collective variables to the case of a strongly coupled plasma.  We mention here the calculation of ``first-principles'' expressions for the free energy over the entire density region of classical plasmas \cite{Kaklyugin,Brilliantov}. In this approach the static structure factor in the Debye-H\"uckel form is employed, strong coupling effects are involved by introducing an upper bound for the collective variables wavevector, as it was done by Debye in the theory of specific heats of solids.

Further approaches are devoted to the study of dynamic properties of coupled Coulomb systems using the representation of the plasma Hamiltonian into collective variables. In a rather incomplete list we mention the approaches in Refs. \cite{Hansen,Suttorp} based on the Mori-Zwanzig theory \cite{Zwanzig,Mori} (or the memory function formalism), the approaches based on the theory of moments \cite{Adamyan}, and the approach based on the quasilocalized charge approximation \cite{Kalman}. 

The present paper is organized as follows. In Sec. \ref{S1}, we apply the approach of Hubbard and Schofield \cite{HSch72} to map the original Hamiltonian of the classical Coulomb system onto a magnetic-like Hamiltonian. The coefficients of the obtained magnetic Hamiltonian are expressed via the ideal gas structure factors, and are calculated exactly. Expanding the anharmonic contributions of the magnetic Hamiltonian we obtain the cluster expansion of a classical plasma.     

The cluster expansion is the starting point for the density expansion of the classical Coulomb system. Explicit calculations are carried out up to the order $n^3$ in Sec. \ref{S2}. Comparison to previous results is carried out briefly.
     
\section{Configurational integral}\label{S1}

Consider a classical plasma consisting of $M$ sorts of ions with $N_a$ ions of a given species $a$ with masses $m_a$ and charges $e_a$. In what follows we will omit the summation bounds in the summation over the particle types $a$ if the summation is carried out from $1$ to $M$. The total number of particles is $N=\sum_{a} N_a$. The system of ions is immersed into a neutralizing background. The plasma system with total volume $V$ and temperature in energy units $T=1/\beta$  is described by the interaction potential

\be \label{U}
U(\vec{r_1}, \ldots ,\vec{r_N})~=~\frac{1}{2} \sum_{a,b} \sum_{i=1}^{N_a} \sum_{j=1}^{N_b} v_{ab}(\vec{r_i},\vec{r_j}) \quad.
\ee 

As far as we consider point-like ions the interaction potential between two particles is given by the Coulomb potential

\be 
 v_{ab}(\vec{r_i},\vec{r_j})~=~\frac{e_a e_b}{|\vec{r_i} -\vec{r_j}|} \quad. \nonumber
\ee

Introduce now the collective variables of the plasma system, the Fourier transforms of the charge density,

\be \label{rho}
\varrho_{\vec{k}}~=~ (\nu)^{-1/2} \sum_a e_a \sum_{i={\cal N}_a}^{{\cal N}_a\,+\, N_a} e^{i \vec{k} \vec{r_i}} \quad ,
\ee
where $\nu = \sum_a e_a^2 N_a$, and ${\cal N}_a~=~\sum_{b=1}^{a-1} N_b$.

The interaction potential can be expressed in terms of the collective variables,
\be \label{Ucoll}
\beta U = \frac{1}{2} \sum_{\vec{k} \not= 0} \alpha(k) \left[ \varrho_{\vec{k}}\varrho_{-\vec{k}} ~-~ 1 \right] \quad,
\ee
with the notation $\alpha(k) \,=\, \kappa^2/k^2$ and $\kappa^2\, =\, (4\pi)/({VT}) \sum_a e_a^2 N_a$ being the square of the inverse Debye radius. The contribution with $\vec{k}=0$ cancels due to the presence of the background.

Our aim now is to calculate the free energy of the plasma system,
\be \label{free}
F ~=~ - T\, \ln \, \, Z_{id}\, Z_c \quad.
\ee 
Here $Z_{id}$ is the ideal part of the partition integral, whereas the configurational integral is given by
\be \label{Zc}
Z_c ~=~ \frac{1}{V^N}\, \int d \vec{r_1} \ldots d \vec{r_N} \, e^{- \beta U(\vec{r_1}, \ldots ,\vec{r_N})} \quad.
\ee
The configurational integral is expressed now through the collective variables \cite{Z54,J58}
\be \label{Zcoll}
Z_c ~=~ \exp \left( \frac{1}{2} \sum_{\vec{k} \not= 0} \alpha(k) \right) \left \langle \exp \left( - \, \frac{1}{2} \sum_{\vec{k} \not= 0} \alpha(k) \, \varrho_{\vec{k}}\varrho_{-\vec{k}} \right) \right \rangle \quad,
\ee  
where the angular brackets denote averaging with respect to an ideal gas system, i.e.,
\be \nonumber
\left \langle \ldots \right \rangle ~=~ \frac{1}{V^N}\, \int d \vec{r_1} \ldots d \vec{r_N} \left( \ldots \right) \quad.
\ee

Using the identity
\be \label{identity}
\exp \left( - \frac{1}{2} a^2x^2 \right) ~=~ \left(2 \pi a^2 \right)^{-1/2} \,\, \int_{-\infty}^{\infty} \exp \left( - \frac{1}{2} \frac{y^2}{a^2} \right) \exp \left(\,-\,ixy \right) \,\, dy 
\ee
one expresses the configurational integral of the interacting system through the configurational integral of a system of noninteracting particles moving in an external field $\varphi_{\vec{k}}$ \cite{HSch72}
\beq \label{Zc1}
Z_c ~&=&~\exp \left( {\frac{1}{2} \sum \alpha(k)} \right) \prod_{\vec{k}} \left \{ \left[ 2 \pi \alpha(k) \right]^{-1/2} \,\, \int d \varphi_{\vec{k}} \right \} \, 
\nonumber \\
&~& \exp \left[ \,-\,\,\frac{1}{2} \left( { \sum \alpha(k)^{-1} \varphi_{\vec{k}}\varphi_{-\vec{k}}} \right)\right]  \left \langle e^{i \sum_{\vec{k}} \varrho_{\vec{k}}\varphi_{-\vec{k}}}\right \rangle  \quad.
\eeq
Note that the field variable $\varphi_{\vec{k}}$ (as well as $\varrho_{\vec{k}}$) is a complex number, $\varphi_{\vec{k}}~=~\varphi_{\vec{k}c}+ i\,\varphi_{\vec{k}s}$ with real $\varphi_{\vec{k}c}$ and $\varphi_{\vec{k}s}$. Therefore in Eq.(\ref{Zc1}) the product of integrals has to be understood in the following manner:
\be  
\int \int d \varphi_{\vec{k}}d \varphi_{-\vec{k}} ~=~\int_{-\infty}^{\infty}\int_{-\infty}^{\infty} d \varphi_{\vec{k}c} d \varphi_{\vec{k}s} \qquad \mbox{and} \qquad \varphi_{\vec{k}}\varphi_{-\vec{k}}~=~\left|\varphi_{\vec{k}c}\right|^2 ~+~\left|\varphi_{\vec{k}s}\right|^2 \quad. \nonumber
\ee
In what follows we will use the shorter notations of Eq.(\ref{Zc1}).

Regard the functional
\be \label{Psi}
\Psi[\varphi] ~=~ \left \langle \exp \left\{ i \sum_{\vec{k}} \varrho_{\vec{k}}\varphi_{-\vec{k}} \right\} \right \rangle \quad .
\ee
The moments of $\varrho_{\vec{k}}$ (i.e., the structure factors of an ideal gas) can be obtained by functional differentiation
\be  \label{moments}
\left \langle \varrho_{\vec{k_1}} \varrho_{\vec{k_2}} \ldots \varrho_{\vec{k_s}} \right \rangle ~=~ (-i)^s \left . \frac{\partial^s \Psi[\varphi]}{\partial \varphi_{\vec{-k_1}}\partial \varphi_{\vec{-k_2}} \ldots \partial \varphi_{\vec{-k_s}}} \right |_{\varphi \equiv 0} \quad .
\ee
Consider further the functional
\be \label{Theta}
\Theta[\varphi] ~=~ \ln \Psi[\varphi] \quad .
\ee
The Taylor expansion of $\Theta$ defines the cumulants of $\varrho$ \cite{K62}
\be \label{cumulants1}
\Theta[\varphi] ~=~ \sum_{n=1}^{\infty}\,\,  \frac{i^n}{n!} \,\, \sum_{\vec{k_1} \ldots \vec{k_n}} \,\, u_n \left( \vec{k_1},\, \vec{k_2},\, \ldots,\, \vec{k_n} \right) \,\,\varphi_{\vec{-k_1}}\varphi_{\vec{-k_2}} \ldots \varphi_{\vec{-k_n}} \quad \, ,
\ee
i.e., the cumulants of order $s$ can be obtained by the functional derivative 
\be \label{cumulants2}
u_s \left( \vec{k_1},\, \vec{k_2},\, \ldots,\, \vec{k_s} \right) ~=~ (-i)^s \left . \frac{\partial^s \Theta[\varphi]}{\partial \varphi_{\vec{-k_1}}\partial \varphi_{\vec{-k_2}} \ldots \partial \varphi_{\vec{-k_s}}} \right |_{\varphi \equiv 0} \quad .
\ee
The first four cumulants read
\beq \label{subcumulants}
u_1\left({\vec{k}}\right)~&=&~\left \langle \varrho_{\vec{k}} \right \rangle ~=~ 0 \quad , \quad \vec{k} ~\neq~ 0 \quad, \nonumber \\
u_2\left(\vec{k_1},\vec{k_2}\right)~&=&~\left \langle \varrho_{\vec{k_1}} \varrho_{\vec{k_2}}\right \rangle \quad , \quad \vec{k_1} ~\neq~ 0 \,,\, \vec{k_2} ~\neq~ 0 \nonumber \\
u_3\left(\vec{k_1},\vec{k_2},\vec{k_3}\right)~&=&~\left \langle \varrho_{\vec{k_1}} \varrho_{\vec{k_2}}\varrho_{\vec{k_3}}\right \rangle \quad , \quad \vec{k_1} ~\neq~ 0 \,,\, \vec{k_2} ~\neq~ 0 \,,\, \vec{k_3} ~\neq~ 0 \quad , \nonumber \\
u_4\left(\vec{k_1},\vec{k_2},\vec{k_3},\vec{k_4}\right)~&=&~\left \langle \varrho_{\vec{k_1}} \varrho_{\vec{k_2}}\varrho_{\vec{k_3}}\varrho_{\vec{k_4}}\right \rangle ~-~ \left \langle \varrho_{\vec{k_1}} \varrho_{\vec{k_2}}\right \rangle \,\, \left \langle \varrho_{\vec{k_3}} \varrho_{\vec{k_4}}\right \rangle \nonumber \\
~&-&~ \left \langle \varrho_{\vec{k_1}} \varrho_{\vec{k_3}}\right \rangle \,\, \left \langle \varrho_{\vec{k_2}} \varrho_{\vec{k_4}}\right \rangle ~-~\left \langle \varrho_{\vec{k_1}} \varrho_{\vec{k_4}}\right \rangle \,\,\left \langle \varrho_{\vec{k_2}} \varrho_{\vec{k_3}}\right \rangle \, , \quad \vec{k_1} ~\neq~ 0  \,,\, \ldots \, ,\, \vec{k_4} ~\neq~ 0\quad .
\eeq
In general one expresses the cumulants of order $s$ $u_s \left( \vec{k_1},\, \vec{k_2},\, \ldots,\, \vec{k_n} \right)$ through the ideal gas structure factor of order $s$ $\left \langle \varrho_{\vec{k_1}} \varrho_{\vec{k_2}} \ldots \varrho_{\vec{k_s}}\right \rangle$ minus all combinations of structure factors of order less than $s$.

Calculate now the ideal gas structure factor of order $n$ ($\vec{k_1} ~\neq~ 0  \,,\, \ldots \, ,\, \vec{k_n} ~\neq~ 0$) \cite{MH61}.
\beq \label{idgassf1}
\left \langle \varrho_{\vec{k_1}} \varrho_{\vec{k_2}} \ldots \varrho_{\vec{k_s}}\right \rangle ~=~ \nu^{-n/2} \frac{1}{V^N} &\,& \int d\vec{r_1} \ldots d\vec{r_N} \, \sum_{a_1,\ldots ,a_n} \,\, e_{a_1} \, e_{a_2} \ldots e_{a_n} \,\, \sum_{i_1={\cal N}_{a_1}+1}^{{\cal N}_{a_1}+N_{a_1}} \ldots \sum_{i_n={\cal N}_{a_n}+1}^{{\cal N}_{a_n}+N_{a_n}} 
\\ \nonumber
&~&\exp \left\{ i \left( \vec{k_1} \vec{r_{i_1}}~+~\ldots~+~\vec{k_n} \vec{r_{i_n}} \right) \right\} \quad .
\eeq 
In the thermodynamic limit (i.e., neglecting terms of the order $1/N$) we obtain for the cumulants
\be \label{cumulantseq}
u_n\left( \vec{k_1},\, \vec{k_2},\, \ldots,\, \vec{k_n} \right) ~=~  N^{-\frac{(n-2)}{2}} \, c_n \, \delta_{\vec{k_1}+\ldots+\vec{k_n}\,,\,0}\,\,, \quad \sum_{i \in {\cal G}} \vec{k_i} ~\neq~ 0 \,\,, \quad {\cal G} \subset \{1,2,\ldots,n\} \quad,
\ee
with $\delta_{ab}$ being the Kronecker delta, and 
\be \label{c_n}
c_n~=~\left( N/\nu \right)^{(n-2)/2}\, \nu^{-1} \, \sum_a e_a^n N_a \quad .
\ee

The conditions in Eq.(\ref{cumulantseq}) reflect the facts that the sum of all $n$ wavevectors being the argument of $u_n\left( \vec{k_1},\, \vec{k_2},\, \ldots,\, \vec{k_n} \right) $ equals zero, whereas all the sums of wavevectors containing less than $n$ wavevectors $\vec{k_i}$ have not to be equal to zero.

Inserting Eq.(\ref{cumulantseq}) into Eq.(\ref{cumulants1}), and using the definition of the functionals $\Theta$ (Eq.(\ref{Theta}) and $\Psi$ (Eq.(\ref{Psi}), we obtain 
\be \label{exp}
\left \langle e^{i \sum_{\vec{k}} \varrho_{\vec{k}}\varphi_{-\vec{k}}}\right \rangle  ~=~ \exp \left( \sum_{n=1}^{\infty}\,\,  \frac{i^n}{n!} \,\, {\sum _{\vec{k_1} \ldots \vec{k_n}}}^\prime \,\,N^{-\frac{(n-2)}{2}} \,c_n\, \delta_{\vec{k_1}+\ldots+\vec{k_n}\,,\,0}  \,\,\varphi_{\vec{-k_1}}\varphi_{\vec{-k_2}} \ldots \varphi_{\vec{-k_n}} \right ) 
\quad \, ,
\ee
where the prime at the summation sign indicates the truncation, i.e., the wavevectors $\vec{k_i}$ are choosen in such a way that all the sums of less than $n$ wavevectors have not to be equal to zero.

Substituting Eq.(\ref{exp}) into Eq.(\ref{Zc1}) we obtain the Hubbard-Schofield representation of the configurational integral for the classical ion mixture \cite{HSch72}
\be \label{H-Sch}
Z_c ~=~\exp \left( {\frac{1}{2} \sum \alpha(k)} \right) \prod_{\vec{k}} \left \{ \left[ 2 \pi \alpha(k) \right]^{-1/2} \,\, \int d \varphi_{\vec{k}} \right \} \, 
~ e^{-\,\tilde{H}[\varphi]} \quad,
\ee
with the effective Hamiltonian
\be \label{effHam}
\tilde{H}[\varphi]~=~\sum_{n=1}^{\infty}{\sum_{\vec{k_1},\ldots,\vec{k_n}}}^{\prime} \,\, w_n\left( \vec{k_1},\, \vec{k_2},\, \ldots,\, \vec{k_n} \right)\,\,\varphi_{\vec{-k_1}}\varphi_{\vec{-k_2}} \ldots \varphi_{\vec{-k_n}} \quad \, ,
\ee
and
\beq \label{w_n}
w_1~&=&~0 \quad ,\nonumber \\
w_2~&=&~\frac{1}{2}\,\,\delta_{\vec{k_1}+\vec{k_2}\,,\,0}\,\, \left[\frac{\vec{k_1}^2}{\kappa^2} ~+~1 \right] \quad , \nonumber \\
w_3~&=&~\frac{i\, c_3}{3!\, N^{\frac{1}{2}}}\,\,\delta_{\vec{k_1}+\vec{k_2}+\vec{k_3}\,,\,0}\,\,  \quad , \nonumber \\
&\vdots & \nonumber \\
w_n~&=&~-\,\,\frac{i^n\, c_n}{n!\, N^{\frac{n-2}{2}}}\,\,\delta_{\vec{k_1}+\vec{k_2}+ \ldots + \vec{k_n}\,,\,0}\,\,  \quad.
\eeq
We rewrite the configurational integral in the form
\be \label{Zc2}
Z_c ~=~{\prod_{\vec{k}}} \left \{ \frac{\exp \left( \alpha(k)/2 \right)}{  \sqrt{ 2 \pi \alpha(k)}}\,\, \int d \varphi_{\vec{k}}  \, 
~ \exp \left[- \frac{1}{2} \left( { \alpha^{-1}(k)\,\,+\,\,1}\right) \varphi_{\vec{k}}\varphi_{-\vec{k}} \right] \right \} 
e^{-\, \sum_{n=3}^{\infty} \, {\cal H}_n}    \quad,
\ee
where the ${\cal H}_n$ are defined by,
\be \label{anharm}
{\cal H}_n ~=~\frac{i^n\, c_n}{n!\, N^{\frac{n-2}{2}}}\,\,{\sum _{\vec{k_1} \ldots \vec{k_n}}}^\prime  \,\, \delta_{\vec{k_1}+\vec{k_2}+ \ldots + \vec{k_n}\,,\,0}\,\,\varphi_{\vec{-k_1}}\varphi_{\vec{-k_2}} \ldots \varphi_{\vec{-k_n}} \quad ,
\ee 
with $c_n$ from Eq.(\ref{c_n}). The ${\cal H}_n$ contain the anharmonic contributions of the auxiliary field variables $\varphi_{\vec{k}}$ into the configurational integral. Neglecting the anharmonic contributions one arrives at the Debye-H{\"u}ckel approximation for the free energy first derived within the framework of collective variables by Zubarev \cite{Z54} (see the next section). Notice that the presence of the Debye potential in the covariance of the Gaussian measure over the $\varphi_{\vec{k}}$'s is the reason for the automatic screening within the present approach. 

In what follows in order to obtain the cluster expansion within the HS approach we have to perform manipulations analogous to those in the Zubarev-Juchnovskij approach. Expanding the exponential $\exp \left\{-\, \sum_{n=3}^{\infty} \, {\cal H}_n \right\}$ one represents the configurational integral as
\beq \label{Zcexpand}
Z_c ~&=~&{\prod_{\vec{k}}} \left \{ \frac{\exp \left( \alpha(k)/2 \right)}{  \sqrt{ 2 \pi \alpha(k)}}\,\, \int d \varphi_{\vec{k}}  \, 
~ \exp \left[- \frac{1}{2} \left( { \alpha^{-1}(k)\,\,+\,\,1}\right) \varphi_{\vec{k}}\varphi_{-\vec{k}} \right] \right \} 
\nonumber \\
&\cdot& 
\left\{ 1 ~+~ {\cal C}_1^\prime ~+~{\cal C}_2^\prime ~+~  {\cal C}_3^\prime ~+~{\cal C}_4^\prime ~+~ \ldots \right\} \quad,
\eeq 
where the abbreviations
\beq \label{S_n}
{\cal C}_1^\prime ~&=&~ {\cal H}_3 ~+~ {\cal H}_4 ~+~ {\cal H}_5 ~+~ \ldots  \quad , \nonumber \\
{\cal C}_2^\prime ~&=&~ \frac{1}{2!} \left\{ \left[ {\cal H}_3 \,\, {\cal H}_3 ~+~ {\cal H}_4 \,\,{\cal H}_4 ~+~  \ldots \,\, \right] ~+~ 2\,\, \left[ {\cal H}_3 \,\, {\cal H}_4 ~+~ {\cal H}_3 \,\, {\cal H}_5 ~+~ \ldots ~+~ {\cal H}_4 \,\, {\cal H}_5 ~+~ \ldots \,\, \right] \right\} \quad , \nonumber \\
{\cal C}_3^\prime ~&=&~ \frac{1}{3!} \left\{ \left[ {\cal H}_3 \,{\cal H}_3\, {\cal H}_3 \,\,+\,\, {\cal H}_4 \,{\cal H}_4\,{\cal H}_4 \,\,+\,\,  \ldots \right] ~+~ 3\,\, \left[ {\cal H}_3 \,{\cal H}_3 \,{\cal H}_4 \,\,+\,\, {\cal H}_3 \,{\cal H}_3 \,{\cal H}_5 \,\,+\,\,~ \right. \right. \ldots 
\nonumber \\
&~&~~~ \left. \left. \,\,+\,\,{\cal H}_4 \,{\cal H}_4 \,{\cal H}_5 \,\,+\,\, \ldots \right] ~+~ 6\,\,\left[ {\cal H}_3 \,\,{\cal H}_4 \,\, {\cal H}_5 ~+~ {\cal H}_3 \,\,{\cal H}_4 \,\, {\cal H}_6\ldots \right] \right\}\quad ,
\eeq
are introduced. Eq.(\ref{Zcexpand}) is an infinite sum of products of integrals of the type
\be \label{type}
\int d \varphi_{\vec{k}}  \, 
~ \exp \left[- \frac{1}{2} \left( { \alpha^{-1}(k)\,\,+\,\,1}\right) \varphi_{\vec{k}}\varphi_{-\vec{k}} \right] \left[\varphi_{\vec{k}} \right]^n \,\, \left[\varphi_{-\vec{k}} \right]^m \quad.
\ee
The result of the integration  over $\varphi_{\vec{k}}$ differs from zero only if  $n=m$ in Eq.(\ref{type}). Having this in mind and taking into account that in ${\cal H}_n$ the sum over the $\vec{k}$'s is truncated, one finds that all linear powers of ${\cal H}_n$ (i.e., ${\cal C}_1^\prime$) and also a lot of products of ${\cal H}_n$ in the higher order ${\cal C}_n^\prime$ ($n\,\,>\,\,1$) vanish under the integration over $\varphi_{\vec{k}}$. In general we have to choose products of
\be \label{products}
w_{n_1}\left( \vec{k}_{11},\,  \ldots,\, \vec{k}_{1n_1} \right)\,\, \ldots \, w_{n_s}\left( \vec{k}_{s1},\, \, \vec{k}_{sn_s} \right)\ldots \,\, \varphi_{-\vec{k}_{11}}\varphi_{-\vec{k}_{12}} \ldots \varphi_{-\vec{k}_{1n_1}} \,\,\ldots\,\,\,\,\varphi_{-\vec{k}_{s1}}\varphi_{-\vec{k}_{s2}} \ldots \varphi_{-\vec{k}_{sn_s}} \quad ,
\ee
(see Eqs.(\ref{effHam},\ref{w_n})) such that one can pair all $\varphi_{\vec{k}_{ij}}$ with other $\varphi_{\vec{k}_{lm}}$ whereby $i \neq l$. 

Consider as an example the pairing of field variables in the lowest nonvanishing product $(1/2)\,{\cal H}_3 \,{\cal H}_3 $ contributing to ${\cal C}_2^\prime$. In this case the integrand is
\beq \label{H3H3}
 &~&{\prod_{\vec{k}}} \left \{ \frac{\exp \left( \alpha(k)/2 \right)}{  \sqrt{ 2 \pi \alpha(k)}}\,\, \int d \varphi_{\vec{k}}  \, 
~ \exp \left[- \frac{1}{2} \left( { \alpha^{-1}(k)\,\,+\,\,1}\right) \varphi_{\vec{k}}\varphi_{-\vec{k}} \right] \right \} 
\nonumber \\
&\cdot& 
\frac{3!}{2} \, \frac{i^3}{3!} \, c_3 N^{- \frac{1}{2}} \,\, \frac{i^3}{3!} c_3 N^{- \frac{1}{2}} {\sum_{\vec{k}_1,\vec{k}_2,\vec{k}_3}}^\prime  \,\,\,  \left( \varphi_{\vec{k}_1}\, \varphi_{\vec{k}_2} \, \varphi_{\vec{k}_3} \right)\,\,\left(\,\varphi_{-\vec{k}_1}\, \varphi_{-\vec{k}_2} \, \varphi_{-\vec{k}_3}\, \right)\,\,\delta_{\vec{k}_1+\vec{k}_2+\vec{k}_3\,,\,0}
\nonumber \\
&=&~{\prod_{\vec{k}}} \left\{ \frac{\exp \left( \alpha(k)/2 \right)}{  \sqrt{ \alpha(k) \,\,+\,\,1}} \, \right\} \,\,  \frac{(-1)^3}{2 \cdot 3!} N^{- {1}} c_3^2\,{\sum_{\vec{k}_1,\vec{k}_2,\vec{k}_3}} \frac{\delta_{\vec{k}_1+\vec{k}_2+\vec{k}_3\,,\,0 }}{\left[\alpha^{-1}(k_1) \,+\,1 \right] \, \left[\alpha^{-1}(k_2) \,+\,1 \right] \, \left[\alpha^{-1}(k_3) \,+\,1 \right]}  
\quad.
\eeq  
It was taken into account here that one has $3!$ possibilities to pair the field variables $\varphi_{\vec{k}}$ in the product ${\cal H}_3 \,{\cal H}_3 $. Notice also that in the r.h.s. of Eq.(\ref{H3H3}) the truncation in the sum over $\vec{k}_1\,,\,\vec{k}_2\,,\,\vec{k}_3$ is omitted. Performing in Eq.(\ref{Zcexpand}) the integration over the field variables $\varphi_{\vec{k}}$ one arrives at the following formula for the configurational integral
\beq \label{Zcexpand2}
Z_c ~&=~&{\prod_{\vec{k}}} \left\{ \frac{\exp \left( \alpha(k)/2 \right)}{  \sqrt{ \alpha(k) \,\,+\,\,1}} \, \right\} \,\,
\left\{ 1 ~+~ {\cal C}_1 ~+~{\cal C}_2 ~+~  {\cal C}_3 ~+~{\cal C}_4 ~+~ \ldots \right\} \quad,
\eeq  
where the ${\cal C}_n$ (without prime) denote the sum of all products of $n$ functions ${\cal H}_i$ ($i\,\,=\,\,3\,\,,\,\,4\,\,,\ldots$) after integration over the field variables. Having in mind the above rules for the pairing of field variables under the integration sign one finds that all the contributions into ${\cal C}_1$ vanish, the second cluster ${\cal C}_2$ is composed of the terms $\left({\cal H}_n \,\,{\cal H}_n \right)$, $n \ge 3$ only, whereas in the third cluster ${\cal C}_3$ only the combinations $\left({\cal H}_n \,{\cal H}_m\,{\cal H}_l \right)$ ($3 \le n \le m \le l$) with $n+m > l$ and $n+m+l$ being an even number remain.  We obtain the following structure for the first three clusters 
\beq \label{S_n2}
{\cal C}_1 ~&=&~0  \quad , \nonumber \\
{\cal C}_2 ~&=&~ \frac{1}{2!} \left\{ \left[ \left({\cal H}_3 \,\, {\cal H}_3\right) ~+~ \left({\cal H}_4 \,\,{\cal H}_4 \right)~+~  \ldots \,\, \right]  \right\} \quad , \nonumber \\
{\cal C}_3 ~&=&~ \frac{1}{3!} \left\{ \left[ \left({\cal H}_4 \,{\cal H}_4\, {\cal H}_4 \right)\,\,+\,\, \left({\cal H}_6 \,{\cal H}_6\,{\cal H}_6 \right)\,\,+\,\,  \ldots \right] ~+~ 3\,\, \left[ \left({\cal H}_3 \,{\cal H}_3 \,{\cal H}_4 \right)\,\,+\,\, \left({\cal H}_4 \,{\cal H}_4 \,{\cal H}_6 \right)\,\,+\,\,~ \right. \right. \ldots 
\nonumber \\
&~&~~~ \left. \left. \,\,+\,\,\left({\cal H}_4 \,{\cal H}_5 \,{\cal H}_5\right) \,\,+\,\, \ldots \right] ~+~ 6\,\,\left[ \left({\cal H}_3 \,\,{\cal H}_4 \,\, {\cal H}_5\right) ~+~ \left({\cal H}_3 \,\,{\cal H}_5 \,\, {\cal H}_6 \right)\ldots \right] \right\}\quad .
\eeq
Here $\left(\ldots\right)$ means that the integration over the field variables $\varphi$ was performed.  

Following Juchnowskij \cite{JG80} we introduce a diagrammatic representation of the expressions appearing in the clusters ${\cal C}_n$. The diagrams consist of lines and points. The connection of lines and points is called a vortic. An unpaired ${\cal H}_n$ contribution occuring in the clusters ${\cal C}_n^\prime$ can be symbolized by a vortic with n open lines

\centerline{
 \psfig{figure=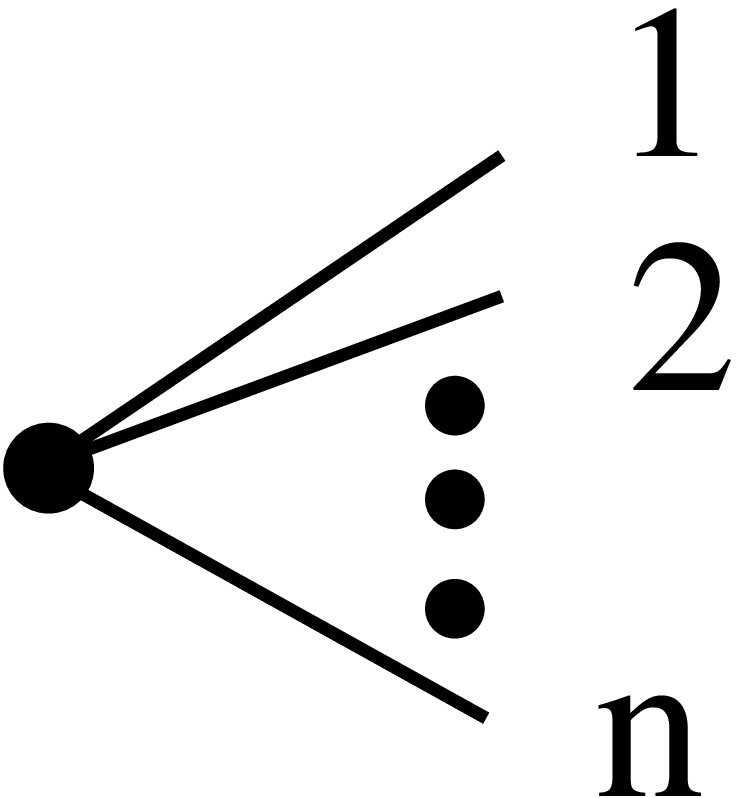,width=1cm,angle=0} \quad ,} 
whereas a paired ${\cal H}_n \, {\cal H}_n$ contributing to the clusters ${\cal C}_n$ can be represented by the following diagram,

\centerline{
 \psfig{figure=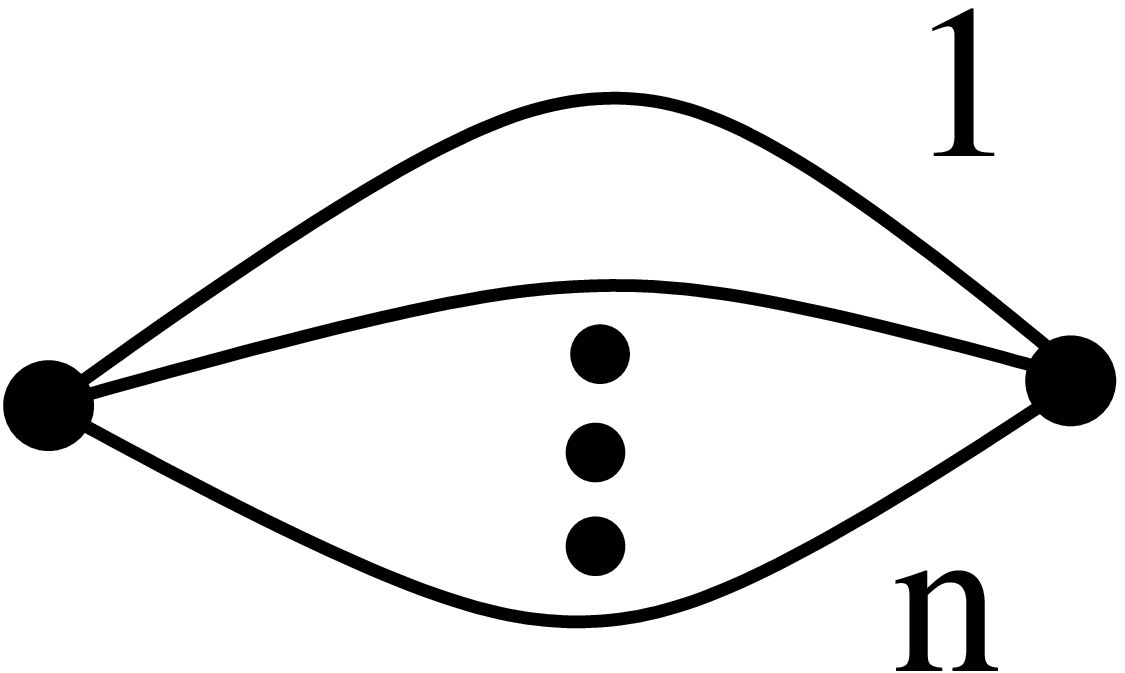,width=2.0cm,angle=0} \qquad .} 

Comparing the diagrammatic representation with the analytic expression from Eq.(\ref{H3H3}) one can establish the following rules for the diagrams:
\begin{enumerate}
\item Outgoing lines are associated with a wavevector $\vec{k}_i$, incoming with $- \, \vec{k}_i$.
\item A vortic with outgoing lines

\centerline{
 \psfig{figure=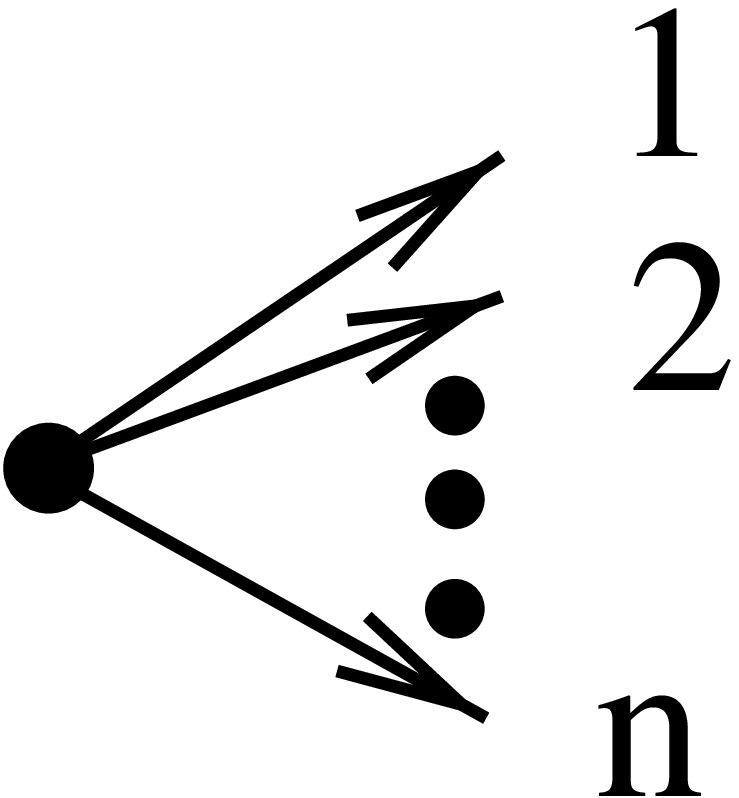,width=1cm,angle=0} \qquad ,} 
 represents the factor
\begin{displaymath}
\frac{i^n\, c_n}{ N^{\frac{n-2}{2}}}\,\,\delta_{\vec{k_1}+\vec{k_2}+ \ldots + \vec{k_n}\,,\,0}\,\,  \quad.
\end{displaymath}
\item n lines connecting two points replace the notations
\begin{displaymath}
\frac{1}{n!} \, \,\frac{1}{\left[\alpha^{-1}(k_1) \,+\,1 \right] \, \left[\alpha^{-1}(k_2) \,+\,1 \right] \, \ldots \, \left[\alpha^{-1}(k_n) \,+\,1 \right]} \quad .
\end{displaymath}
\item One has to count the number of topological identical diagrams which differ only by the numeration of the points. (In what follows this number will be shown explicitely.)
\item Summation over the wavevectors has to be performed. 
\end{enumerate}

With the thus defined diagrams one easily expresses the clusters ${\cal C}_n$. From Eq.(\ref{S_n2}) and calculations analogous to that in Eq.(\ref{H3H3}) one finds that the cluster ${\cal C}_n$ is given by the sum of all possible diagrams satisfying the following conditions:
\begin{enumerate}
\item The diagrams consist of $n$ points and closed lines.
\item The minimal number of outgoing and incoming lines in each point of a corresponding diagram equals three.
\item Selftruncation is forbidden, i.e., closed loops are absent, each line connects two different points of a diagram. In addition a connected diagram or a connected subdiagram of a disconnected diagram should have the following property. By cutting an arbitrary number of connections of a certain (but arbitrary) point with other points but keeping at least one connection one remains with a connected diagram (or subdiagram) still. (This rule represents the prime in Eq.(\ref{anharm}).) For example due to this rule the following diagrams should be excluded:
$$\begin{minipage}[htbp]{1.4cm} \centerline{\psfig{figure=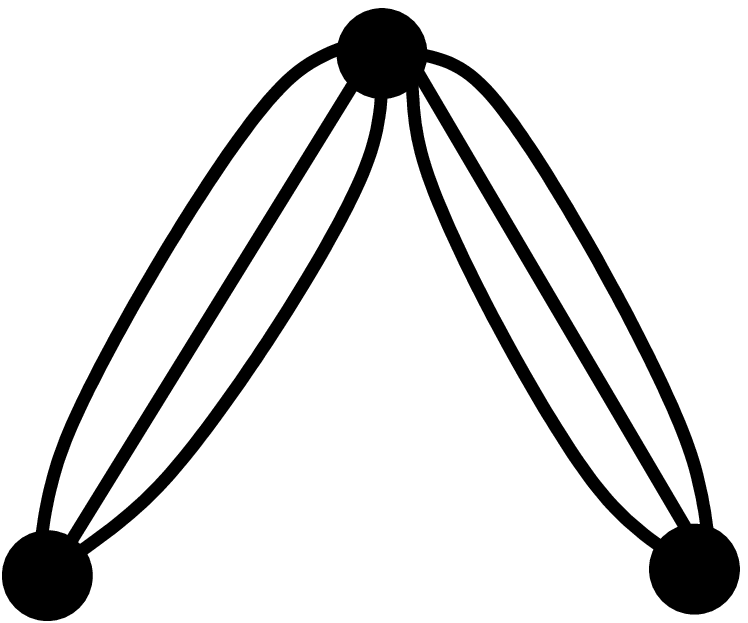,width=1.1cm,angle=0}}\end{minipage} \qquad \begin{minipage}[htbp]{1.4cm} \centerline{\psfig{figure=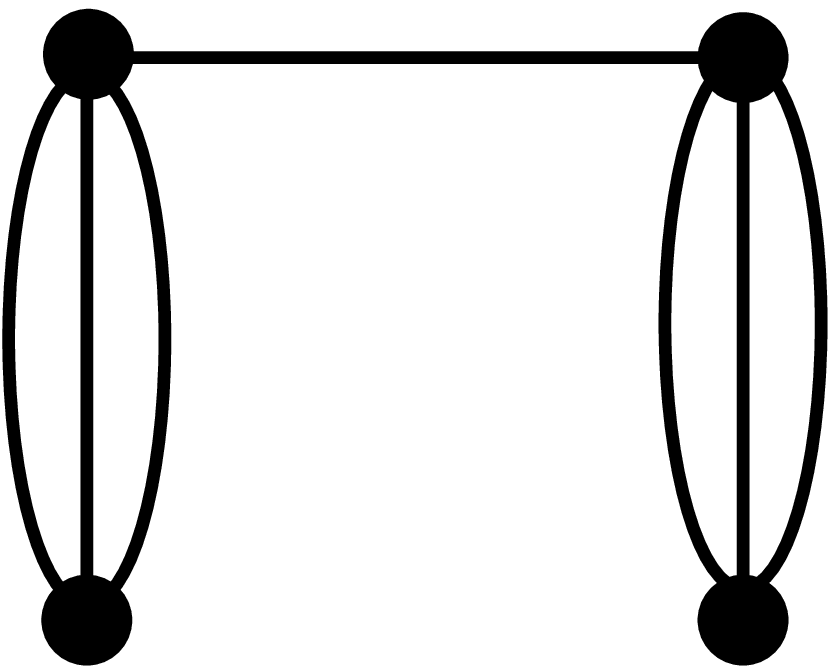,width=1.1cm,angle=0}}\end{minipage} \quad .$$ 

Since our final aim is the calculation of the free energy we rewrite Eq.(\ref{Zcexpand2}) in a more convenient form,
\beq \label{Zcexpand3}
Z_c ~&=~&{\prod_{\vec{k}}} \left\{ \frac{\exp \left( \alpha(k)/2 \right)}{  \sqrt{ \alpha(k) \,\,+\,\,1}} \, \right\} \,\,
\exp \left\{ {\cal S}_2 ~+~  {\cal S}_3 ~+~{\cal S}_4 ~+~ \ldots \right\} \quad,
\eeq 

The cluster ${\cal S}_n$ can be expressed by the same diagrams as the cluster ${\cal C}_n$ with one additional condition:
\item From all the diagrams contributing to ${\cal S}_n$ we have to choose only irreducible diagrams, i.e., those which do not represent a combination of lower order diagrams. (The latter arise in Eq.(\ref{Zcexpand2}) by expanding the exponent in Eq.(\ref{Zcexpand3})). For example the diagram 
$$\begin{minipage}[htbp]{1.4cm} \centerline{\psfig{figure=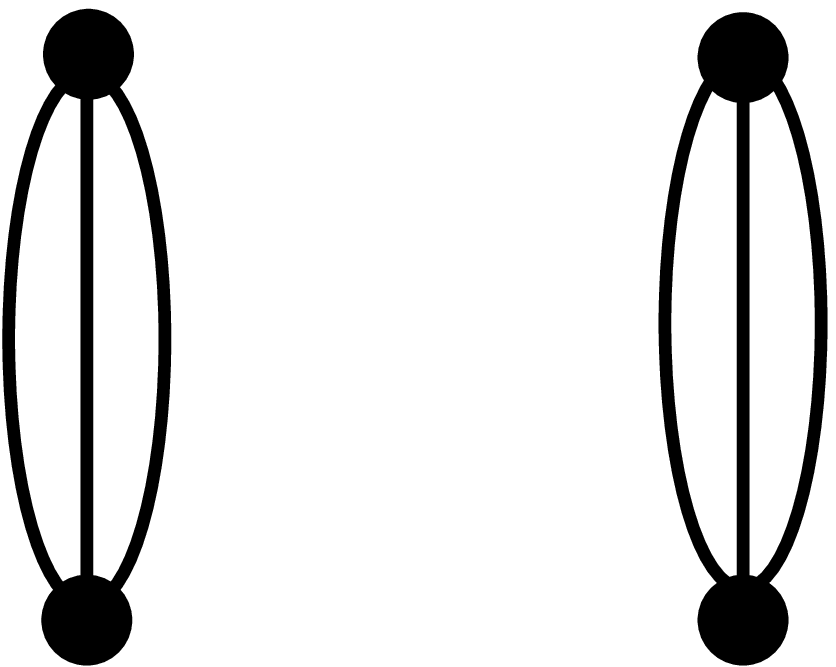,width=1.1cm,angle=0}}\end{minipage}$$
contributes to ${\cal C}_4$ but does not contribute to ${\cal S}_4$.
\end{enumerate}
 
The first cluster integrals ${\cal S}_n$ will be considered in the next section.

\section{virial expansion} \label{S2}
 Our aim now is to obtain the virial expansion for the free energy of the ion mixture. The basic expression that is used as a starting-point is the cluster expansion for the interaction part $F_{\mathrm{c}}$ of the Helmholtz free energy per volume $V$ in units of $T=1/\beta$,
\be \label{F_c}
{\beta F_{\mathrm{c}}\over V}=-\,\Biggl(f_{DH}
+\,\, \frac{1}{V} \, \sum_{n=2}^{\infty}{\cal{S}}_{n} \Biggr)\,,
\ee
where
\be \label{F_DH}
f_{DH}~=~ -\,   \frac{1}{2 V} {\sum_{\vec{k}}}  \,\, \left\{ \,\, \alpha(k)  ~-~  \ln \left[ \alpha(k) \,\,+\,\,1 \, \right]\right\} ~=~ {\kappa^{3}\over 12\pi}\,  \quad,
\ee
is the leading Debye-H\"uckel contribution to the free energy \cite{Z54}.

In the following part of this section we will consider the first cluster integrals and extract the leading terms of the density expansion within each cluster integral.
\subsection{Second cluster integral}
The second cluster integral is given by the sums of diagrams
\be \label{second}
{\cal{S}}_{2}~=~ \frac{1}{2} \left( \hspace{0.2cm} \begin{minipage} [htbp]{5cm} \centerline{\psfig{figure=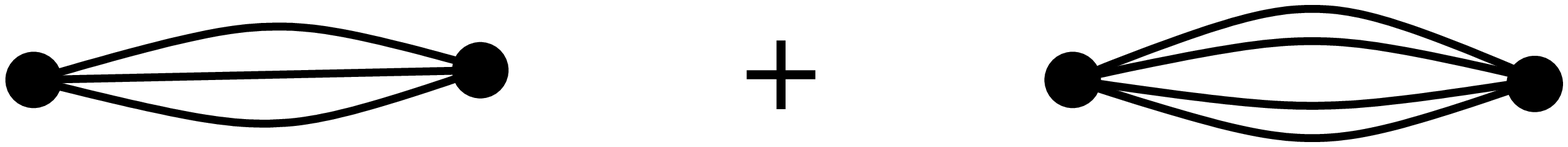,width=4.4cm,angle=0}}\end{minipage} \hspace{0.2cm} + \,\, \ldots \,\, \right)\, ,
\ee
or in analytic representation
\be \label{second2}
{\cal{S}}_{2}~=~\frac{1}{2} \,\, \sum_{n=3}^{\infty} \frac{(-1)^n c_n^2}{n! N^{n-2}} \,{\int \frac{d \vec{k}_1 \, d \vec{k}_2 \, \ldots \, d \vec{k}_n}{\left(2\pi\right){3n}}} \frac{\delta_{\vec{k}_1+\vec{k}_2+\,\ldots \, + \vec{k}_n\,,\,0 }}{\left[\frac{k_1^2}{\kappa^2} \,+\,1 \right] \, \left[\frac{k_2^2}{\kappa^2} \, +\,1 \right] \, \ldots \, \left[\frac{k_n^2}{\kappa^2} \,+\,1 \right]}  
\quad.
\ee
Substituting the $c_n$ values from Eq.(\ref{c_n}), and using the convolution theorem and the expression for the inverse Fourier transformation of 
\be \label{inverse}
\int  \frac{d\,\vec{k}}{\left(2\pi\right)^3} \frac{4 \pi l_{ab}}{k^2 \,\,+\,\, \kappa^2}\,\, e^{i \vec{k} \vec{r}} ~=~ g_{ab}(r) ~\equiv   ~ \frac{l_{ab}}{r}\,e^{- \kappa r} \quad , \qquad
l_{ab}=e_a e_b \beta \,\,,
\ee
we find the familiar expression for the second cluster integral 
\beq \label{second3}
{\cal{S}}_{2}~&=&~ \sum_{a,b} \frac{N_a N_b}{2 V}\,\, \sum_{n=3}^\infty \, \frac{(-1)^n}{n!}\,\int\, d \vec{r} \, \left[\, g_{ab}(r)\,\right]^n
\nonumber \\
~&=&~\sum_{a,b} \frac{N_a N_b}{2 V}\,\, \,\int\, d \vec{r} \,\left[\Phi_{ab}(r)\, \,-\,\,\frac{1}{2} \, g_{ab}^2(r) \, \right] \, ,
\eeq
where the generalized Mayer function
\beq \label{Mayer}
\Phi_{ab}\;=\;e^{-\,g_{ab}(r)}\,\,-\,\,1\,\,+\,\,g_{ab}(r)\,\, ,
\eeq
is introduced.

Thus performing the inverse Fourier transform one can establish the following rules of evaluation of a cluster expansion graph in the coordinate representation:
\begin{enumerate}
\item Each point is associated with a particle of species $a$, and provides an integration over the coordinate space - $n \, c_a \, \int \, d \vec{r}_a$, the total particle density $n=N/V$ and the particle concentration of species $a$, $c_a=N_a/N$, are introduced here. 
\item $l$ solid lines $\begin{minipage}[htbp]{1.3cm} \centerline{\psfig{figure=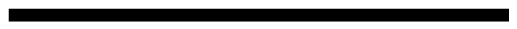,width=1.1cm,angle=0}}\end{minipage}$ connecting the points $a$ and $b$ give a factor $((-1)^l/l!) \, [g_{ab}(r_{ab})]^l$ where $r_{ab}\,\,=\,\,|\vec{r}_a - \vec{r}_b|$.
\item Introducing a graphic representation of the generalized Mayer function,
\be \label{Mayergraph}
\begin{minipage}[htbp]{1.8cm} \centerline{\psfig{figure=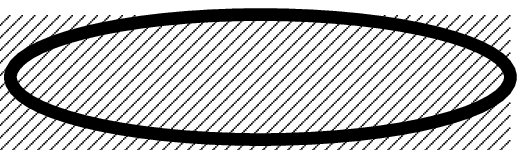,width=1.6cm,angle=0}}\end{minipage}~=~\begin{minipage}[htbp]{1.3cm} \centerline{\psfig{figure=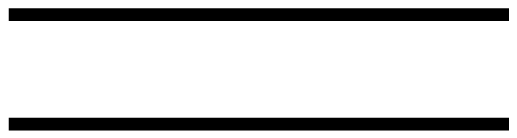,width=1.1cm,angle=0}}\end{minipage}~+~\begin{minipage}[htbp]{1.3cm} \centerline{\psfig{figure=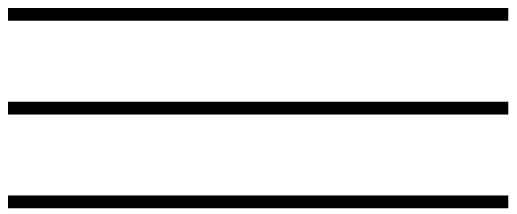,width=1.1cm,angle=0}}\end{minipage}\,\,  + \,\, \ldots \,\, ,
\ee
one associates a bond $\begin{minipage}[htbp]{1.8cm} \centerline{\psfig{figure=figures/mayer.eps,width=1.6cm,angle=0}}\end{minipage}$ connecting the points $a$ and $b$ with the Mayer function $\Phi_{ab}(r_{ab})\;=\;\exp[-\,g_{ab}(r_{ab})]\,\,-\,\,1\,\,+\,\,g_{ab}(r_{ab})$.
\item Finally, summation over the particle types has to be performed.
\end{enumerate}

Given this rules one represents the second cluster integral in the following manner
\be \label{Second}
{\cal{S}}_{2}~=~ \frac{1}{2} \left( \hspace{0.2cm} \begin{minipage} [htbp]{6cm} \centerline{\psfig{figure=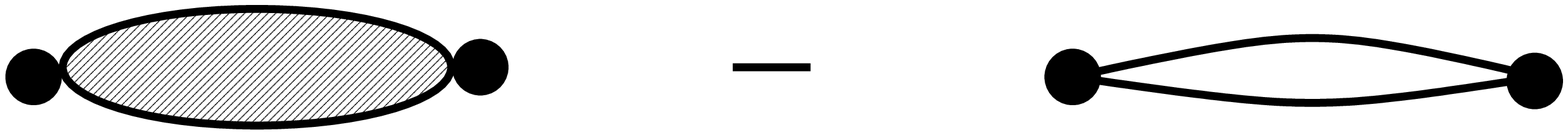,width=5.4cm,angle=0}}\end{minipage} \right)\, .
\ee

In order to get the density expansion of the second cluster integral we expand the corresponding expression in Eq.(\ref{second3}) in powers of $\kappa$. Up to the order $n^{3}$ we get,
\beq \label{second-density}
{\cal{S}}_{2}&=&\,2\pi\sum_{a,b}\frac{N_{a}N_{b}}{V} \,  \int\limits_{0}^{\infty}\!\!drr^{2}\biggl\{ \,\, \biggl[\,-{1\over 6}\biggl({l_{ab}\over r}\biggr)^{\!3} \left(e^{-3\kappa r}-1\right) + f_{ab}(r) \,\biggr]
+\biggl[{1\over {4!}}\biggl({l_{ab}\over r}\biggr)^{\!4} \left(e^{-4\kappa r}-1+4\kappa r \right) + l_{ab}\kappa f_{ab}(r) \,\biggr]\\[1.5ex]
&+&\biggl[-{1\over {5!}}\biggl({l_{ab}\over r}\biggr)^{\!5} \left(e^{-5\kappa r}-1+5\kappa r - \frac{25}{2} \kappa r \right) + {1\over 2}(l_{ab}\kappa)^{2} f_{ab}(r) -\,{1\over 2}\,l_{ab}\kappa^{2} r \left( f_{ab}(r) + \frac{1}{6}\biggl({l_{ab}\over r}\biggr)^{\!3} \right)\,\biggr] \biggr\}
\,,\nonumber
\eeq
where the truncated bond
\be \label{trunc-bond}
f_{ab}(r)=\exp\left(-\,\frac{l_{ab}}{r}\right)-1+{l_{ab}\over r}-{1\over 2}
\biggl({l_{ab}\over r}\biggr)^{\!2} \,\,,
\ee
has been introduced.

Similar calculations of the second cluster integral at the order $n^{3}$ for a quantum plasma have been performed by Kahlbaum \cite{K97} and Ortner \cite{O97}. 

A straightforward calculation of the integrals in Eq.(\ref{second-density}) leads to the following result,
\beq \label{second-final}
{\cal{S}}_{2}&=&\,V\,\biggl\{ n^2\sum_{a,b}c_ac_b l_{ab}^3\left[\frac{\pi}{3}\ln{\kappa\l_{ab}}\,\,+\,\,a_2\right]~+~n^2 \kappa \sum_{a,b}c_ac_bl_{ab}^4 \left[\frac{\pi}{3}\ln{\kappa\l_{ab}}\,\,+\,\,b_2\right]
\nonumber\\
&+& n^2 \kappa^2 \sum_{a,b}c_ac_b l_{ab}^5 \left[\frac{5\pi}{24}\ln{\kappa\l_{ab}}\,\,+\,\,c_2\right] \biggr\} \quad,
\eeq
with the numerical constants
\beq \label{constants}
a_2=\frac{\pi}{3}\left[2C_E-\frac{11}{6}+\ln{3}\right]= 0.439519\dots\,,
\nonumber\\
b_2=\frac{\pi}{3}\left[2C_E-\frac{17}{6}+\ln{4}\right]= - \, 0.30641\dots\,,
\nonumber\\
c_2=\frac{5\pi}{24}\left[2C_E-\frac{203}{60}+\ln{5}\right]= - \, 0.4054\dots\,,
\eeq
$C_{E}$ being the Euler-Mascheroni constant, $C_{E}=0.57721566\dots
$.

Thus we recover the result of Ab{\'e} \cite{A59} (i.e., the density expansion of the free energy up to the order $n^2$) and have written down all terms of the order $n^3$ and $n^3\ln n$ stemming from the second cluster integral.


\subsection{Third cluster integral} \label{thirdsection}
Following the general rules of construction of the cluster integrals of arbitrary order one expresses the third cluster integral through the following sum of diagrams \cite{JG80}
\beq \label{third}
{\cal{S}}_{3}~=~ \frac{1}{3!} \,\,\left\{ \,\, 3 \,\, \left[ \,\, \begin{minipage} [htbp]{1.3cm} \centerline{\psfig{figure=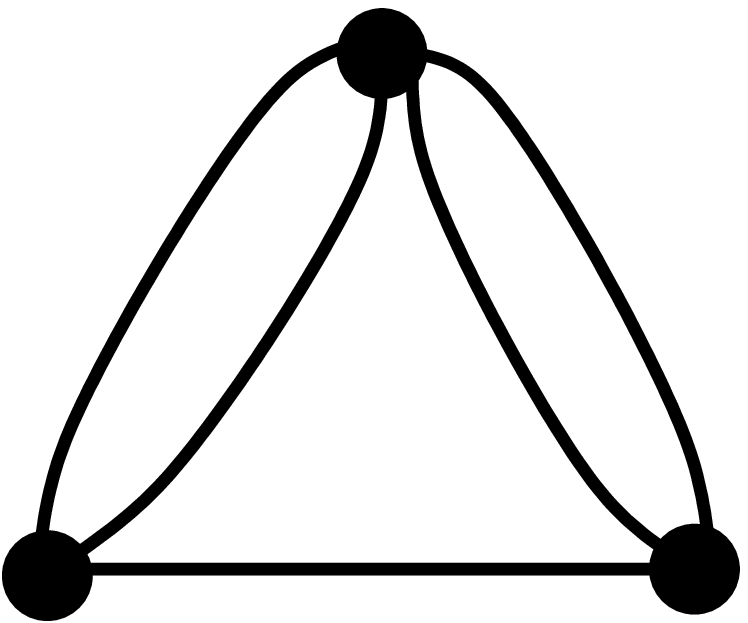,width=1.1cm,angle=0}}\end{minipage}~ + ~ 2 \,\, \begin{minipage} [htbp]{1.3cm} \centerline{\psfig{figure=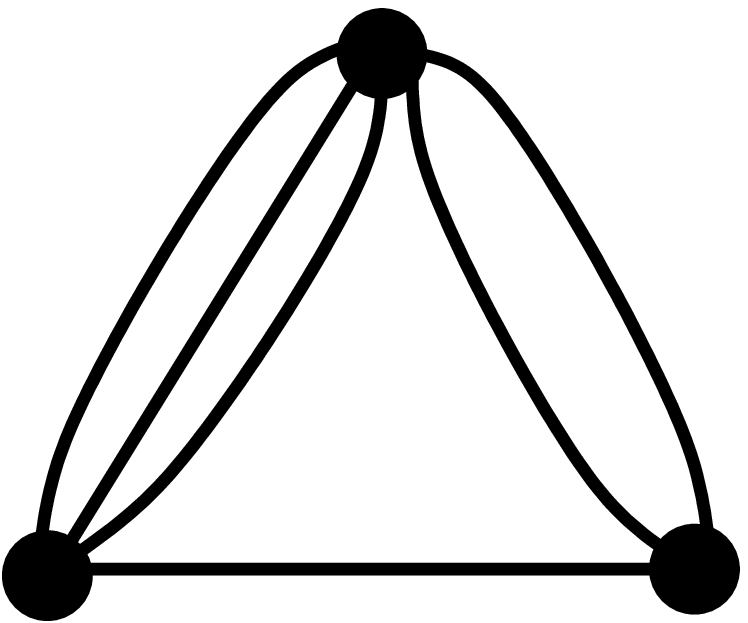,width=1.1cm,angle=0}}\end{minipage}\,\, + \,\, \ldots \,\, \right] 
~ + ~ \left[ \,\,   \begin{minipage} [htbp]{1.3cm} \centerline{\psfig{figure=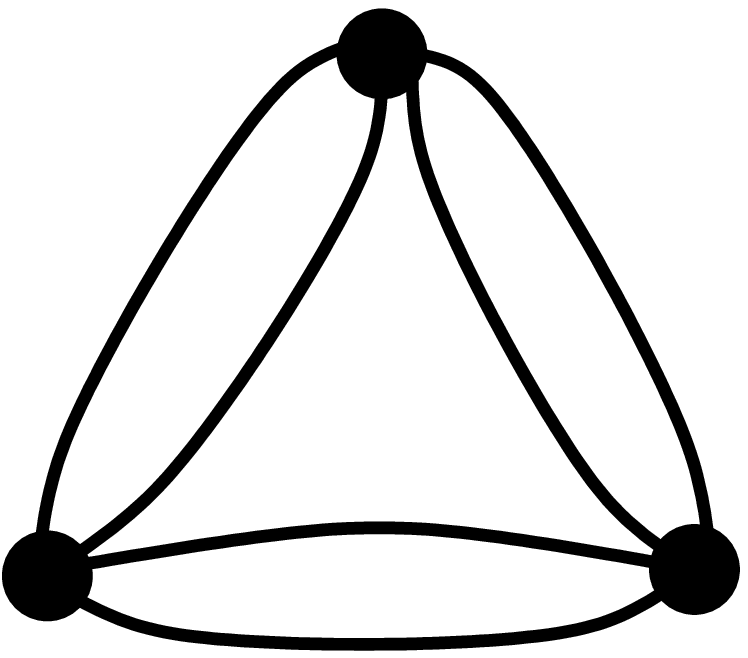,width=1.1cm,angle=0}}\end{minipage} ~ + ~ 2 \,\, \begin{minipage} [htbp]{1.3cm} \centerline{\psfig{figure=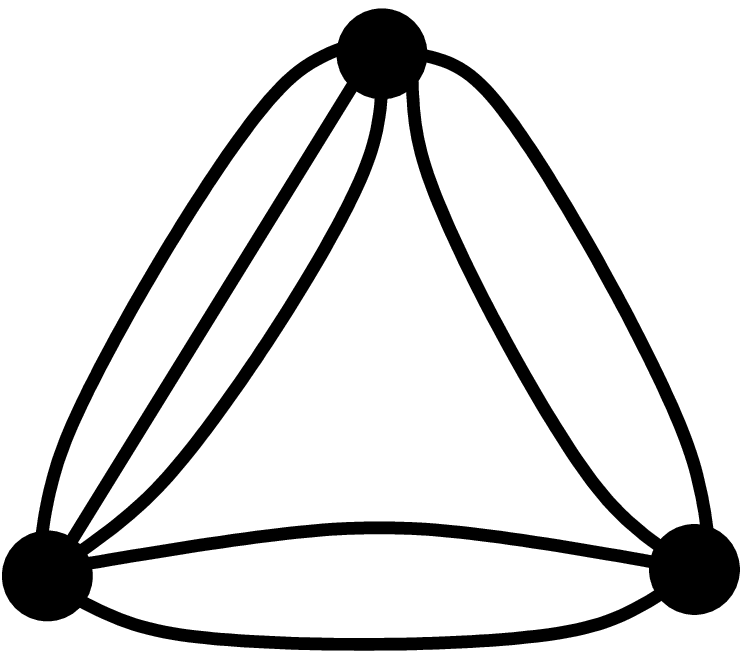,width=1.1cm,angle=0}}\end{minipage}\,\,  + \,\, \ldots \,\, \right] \, \right\} \, .
\eeq
With the graphic representation of the generalized Mayer function we represent the third cluster integral in the compact manner \cite{J58,F62,Sch68},
\beq \label{Third}
{\cal{S}}_{3}~=~ \frac{1}{3!} \,\,\left\{ \,\, 3 \,\,  \begin{minipage} [htbp]{2.3cm} \centerline{\psfig{figure=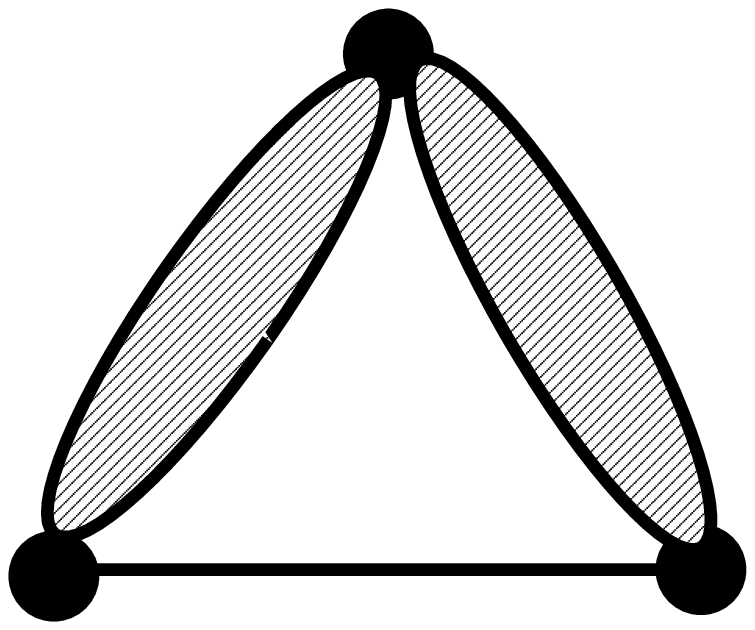,width=2.1cm,angle=0}}\end{minipage}~ + ~  \begin{minipage} [htbp]{2.3cm} \centerline{\psfig{figure=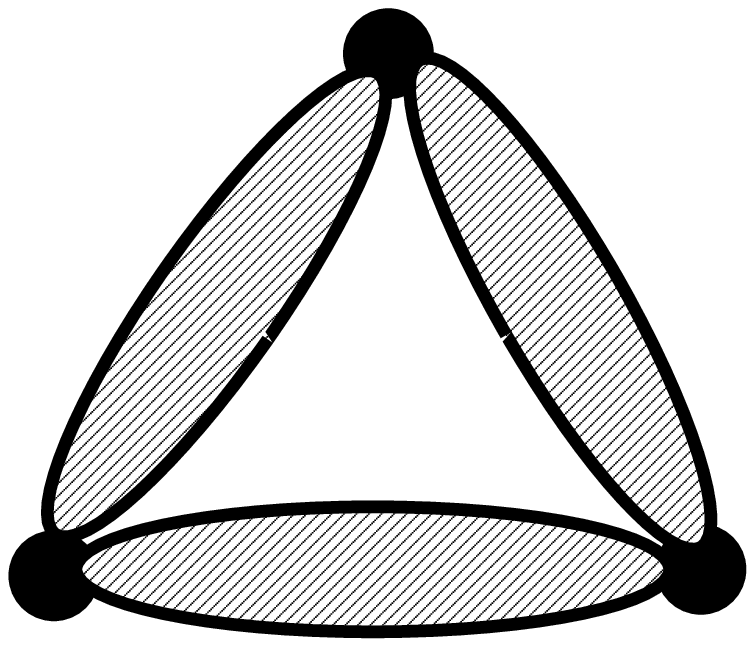,width=2.1cm,angle=0}}\end{minipage}  \, \right\} \, .
\eeq

Using the diagrammatic rules one easily writes down the explicit expression for the third cluster integral
\be
\label{eq:C3}
{\cal S}_3={n^3\over 6} \sum_{a,b,c} c_{a}c_{b}c_{c} \int\!\!d\vec{r}_{a}d\vec{r}_{b}
d\vec{r}_{c}\Bigl[\,-\,3g_{ab}(r_{ab})\Phi_{ac}(r_{ac})\Phi_{bc}(r_{bc})~+~\Phi_{ab}(r_{ab})\Phi_{ac}(r_{ac})\Phi_{bc}(r_{bc})
\Bigr]\,.
\ee

However, we are interested in the density expansion of the free energy. The third cluster integral like all cluster integrals is a function of density $n$. We formulate now some general rules of picking out the leading terms of a density expansion of a certain cluster integral. A graph made of $m$ points and $l$ solid lines (i.e., it is a contribution to the $m$th cluster integral) is proportional to $n^m \int d \vec{r}_1 \ldots d \vec{r}_m \, g^l$. Here the occurence of alltogether $l$ Debye functions $g$ with different arguments is scetched only schematically. Substituting $\vec{x}_i=\kappa \vec{r}_i$ one finds that the above graph gives a density contribution to the free energy density (with the additional factor $1 \over V$, see Eq.(\ref{F_c})) of the order
$$
n^m \kappa^{3} \kappa^{-3m} \kappa^{l} ~\sim ~ n^{3/2\,+\,l/2\,-\,m/2}~.
$$
It should be mentioned that this simple estimate is valid only as far as the remaining dimensionless integral is convergent. Especially, it is not correct when bonds $g^p(r)$ of one argument $r$ with $p \ge 3$ are present. These bonds are nonintegrable due to the singularity at short distances $r$. However, these short range divergencies are spurious. They are just an artefact of the expansion of Boltzmann factors in the Mayer-like graphs in powers of the screened or unscreened Coulomb potential. Since all Boltzmann factors are integrable at short distances it is always possible to eliminate these unphysical divergencies at any order by a suitable collection of the dangerous contributions \cite{Lavaud,referee}. For example the $g^p(r)$ bonds ($p \ge 3$) may be resummed into the bond $\Psi - g^2/2$. At low densities, this bond can be replaced by the bare truncated bond $f(r)$ (see Eq.(\ref{trunc-bond})). This is an integrable bond except for the logarithmic divergency at large distances. The latter divergency can be finally removed using the bond 
$$-{1\over 6}\biggl({l_{ab}\over r}\biggr)^{\!3} \left(e^{-3\kappa r}-1\right) + f_{ab}(r) \quad ,$$
which leads to the appearance of the logarithmic term $\ln \kappa$. This is just the content of the renormalization procedure employed in Eq.(\ref{second-density}). It will be used in the remaining of this section. Nevertheless, the above estimation of the order of a cluster integral is correct as far as the leading contributions of the $m$th ($m \ge 3$) cluster integral are looked for. These contributions are obtained with bonds $g$ and $g^2$ only for which the scaling analysis is applicable. One also proofs that the minimal number $l_{min}(m)$ of solid lines of a graph contributing to the mth cluster integral is $2m - [m/2]$ with $[\lambda]$ being the integer number of $\lambda$. Therefore the leading contribution of the $m$th cluster integral is of the order of
$$
\frac{1}{2} \left(\, 3+m-\,\left[ \frac{m}{2} \right] \, \right) \,\, .
$$
The leading contribution from the third cluster integral is the graph with 5 solid lines (stemming from the ${\cal H}_3 \,{\cal H}_3 \,{\cal H}_4 $ term, see Eq.(\ref{S_n2})),
$$
{\cal{S}}_{3}^l~=~\frac{1}{3!}  \,\, 3 \,\,  \begin{minipage} [htbp]{1.3cm} \centerline{\psfig{figure=figures/third1.eps,width=1.1cm,angle=0}}\end{minipage} ~=~-~ {n^3 \over 2} \sum_{a,b,c} c_{a}c_{b}c_{c} \int\!\!d\vec{r}_{a}d\vec{r}_{b}
d\vec{r}_{c} \,\, g_{ab}(r_{ab}) \, \frac{1}{2} \, \left[ \, g_{ac}(r_{ac})\, \right]^2 \, \frac{1}{2} \, \left[ \, g_{bc}(r_{bc})\, \right]^2 \quad ,
$$
and gives a contribution to the order $n^{5/2}$ \cite{Haga},
\begin{equation} \label{S3l}
{\cal{S}}_{3}^l~=~-~{b_{3}} \frac{V n^3}{\kappa} \sum_{a,b,c}c_{a}c_{b}c_{c}l_{ab}l_{ac}^{2}
l_{bc}^{2} \,\,.
\end{equation}
The numerical constant $b_3$ (an integral representation was already given by Haga \cite{Haga}) is calculated in Appendix \ref{AppA}. It is expressable through the Euler's dilogarithm $\mathrm{Li}_{2}(x)=\sum_{k=1}^{\infty}x^{k}/k^{2}$, $|x|\le 1$ and equals in accordance with \cite{K97}
\be \label{a_1}
b_{3}=2\pi^{2}\biggl[{\pi^{2}\over 12} + \mathrm{Li}_{2}\biggl(-{1\over 3}\biggr)\biggr]
=10.13477\dots\,.
\ee
Following the general scheme the $n^3$ terms should be obtained from all graphs with 6 solid lines, i.e., from the diagrams
\be
\begin{minipage} [htbp]{1.3cm} \centerline{\psfig{figure=figures/third3.eps,width=1.1cm,angle=0}}\end{minipage} ~ \mathrm{and} ~ \begin{minipage} [htbp]{1.3cm} \centerline{\psfig{figure=figures/third2.eps,width=1.1cm,angle=0}}\end{minipage} \quad ,
\ee
representing the ${\cal H}_4 \,{\cal H}_4 \,{\cal H}_4 $ and ${\cal H}_3 \,{\cal H}_4 \,{\cal H}_5 $ terms.

However, the corresponding integrals are divergent. The divergencies occur at short distances and are connected with a singular accumulation of $g^2$ bonds or with the occurence of a bond $g^3$. These short range divergencies are unphysical and can be removed by a renormalization procedure as discussed above. Regard therefore the two corrections to the leading contribution ${\cal{S}}_{3}^l$,
\be \label{delta1}
\delta_1{\cal S}_3={n^3\over 6} \sum_{a,b,c} c_{a}c_{b}c_{c} \int\!\!d\vec{r}_{a}d\vec{r}_{b}
d\vec{r}_{c} \,\, \Phi_{ab}(r_{ab})\Phi_{ac}(r_{ac})\Phi_{bc}(r_{bc})
\,\,.
\ee
and 
\be
\label{delta2}
\delta_2{\cal S}_3=~-~{n^3\over 2} \sum_{a,b,c} c_{a}c_{b}c_{c} \int\!\!d\vec{r}_{a}d\vec{r}_{b}
d\vec{r}_{c} \, g_{ab}(r_{ab}) \, 2 \, \left[ \Phi_{ac}(r_{ac})\,\, - \,\, \frac{1}{2} g_{ac}^2(r_{ac}) \,\, \right] \, \Phi_{bc}(r_{bc})\,.
\ee
and expand the corresponding contributions in terms of $\kappa$. 

First, consider the term $\delta_1{\cal S}_3$. As it is shown in Appendix \ref{AppB} we get up  to the order $n^3$,
\beq
\label{delta1a}
\delta_1{\cal S}_3= -~\,{V n^3} \sum_{a,b,c} c_{a}c_{b}c_{c}l_{ab}^2l_{ac}^{2}
l_{bc}^{2}\left[\,\,\frac{\pi^4}{12} \ln{\kappa l_{ac}}~+~c_{3\alpha}\,\,\right]\,\,,
\eeq
where the constant
\be \label{c_3alpha}
c_{3\alpha}~=~\frac{\pi^4}{12} \left[ \,\, \lim_{\delta \to 0} \,\, 16 \pi \,\,\int_0^{1/\delta}\frac{dk}{k} \, \arctan^3{\frac{k}{2}}~+~2 \pi^4\ln{\delta}~+~C_E~-~\frac{3}{2}\,\,\right]~=~-~2.89476\dots \,\,,
\ee
has been introduced.
Consider now the term $\delta_2{\cal S}_3$. Up to the order $n^3$ we get (see Appendix \ref{AppB})
\be
\label{delta2b}
\delta_2{\cal S}_3=~\,{V n^3} \sum_{a,b,c} c_{a}c_{b}c_{c}l_{ab}l_{ac}^{3}l_{bc}^{2}\,\,\left[\,\, \frac{2 \pi^2}{3} \ln^2{\kappa l_{ac}} + \frac{8 \pi^2}{3} \left( \, C_E - \frac{17}{12}+ \frac{1}{2}\ln{3}\,\right) \ln{\kappa l_{ac}} ~+~ c_{3\beta} \,\right]\,\,,
\ee
with
\beq
\label{c_3beta}
 c_{3\beta}~=&~&\frac{2 \pi^2}{3} \left\{ \,\, 2\,\,\Biggl[C_E - 1+\ln{3}\,\Biggr]\,\,\Biggl[C_E - \frac{11}{6}+\ln{3}\,\Biggr]~+~ \ln{3} \,\left[\,8\ln{2}-3\ln{3}\,\right]  - 8\ln^2{2} - \frac{\pi^2}{3} \right.
\nonumber\\
&~& + \left.\frac{121}{18}~+~2 C_E^2~-~\frac{17}{3}C_E~-~4  \mathrm{Li}_{2}\biggl(-{1\over 4}\biggr) - 2  \mathrm{Li}_{2}\biggl(-{2\over 3}\biggr)\,\,\right\}~=~ 8.85348 \dots \,\,.
\eeq
Notice that the leading term of $\delta_2{\cal S}_3$ is of the order $n^3\ln^2{n}$. 
\subsection{Fourth cluster integral} \label{fourthsection}
The diagrammatic representation of the fourth cluster integral is found from the above diagrammatic rules and reads in accordance with \cite{JG80},
\beq \label{fourth}
{\cal{S}}_{4}~&=&~ \frac{1}{4!} \,\,\left\{ \,\, 6 \,\, \left[ \,\, \begin{minipage} [htbp]{1.3cm} \centerline{\psfig{figure=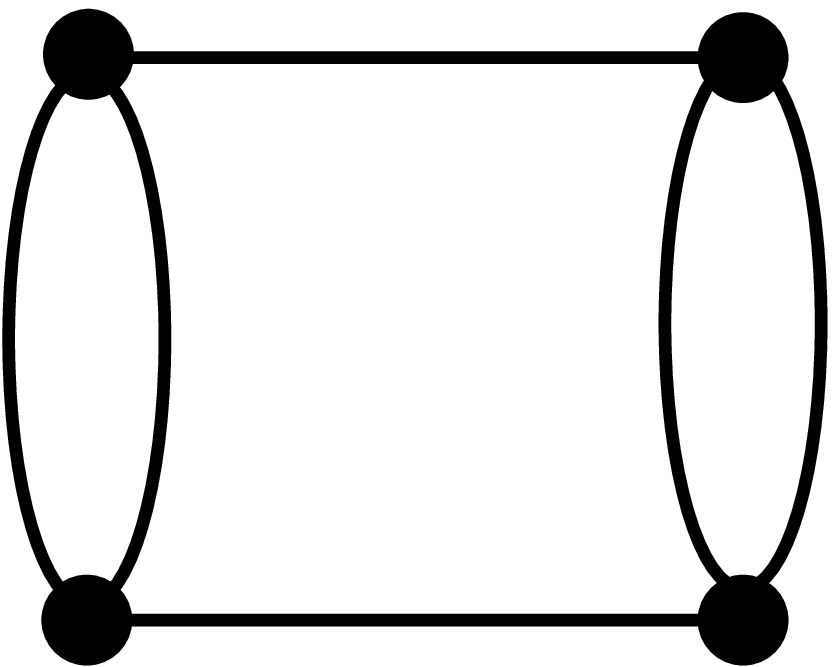,width=1.1cm,angle=0}}\end{minipage}~ + ~  \ldots \,\, \right] 
~ + ~ \left[ \,\,   \begin{minipage} [htbp]{1.3cm} \centerline{\psfig{figure=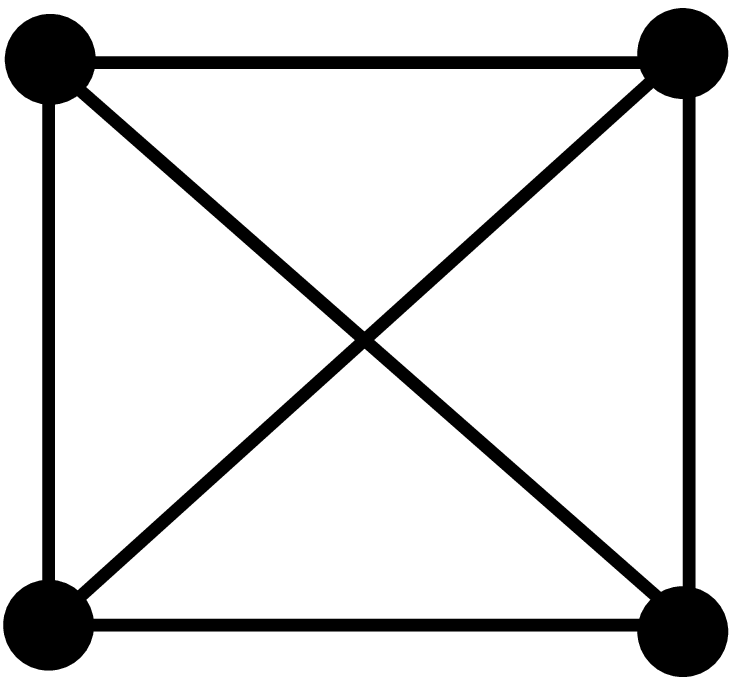,width=1.1cm,angle=0}}\end{minipage} ~ + ~  \ldots \,\, \right] 
~ + ~12 \,\,  \left[ \,\,   \begin{minipage} [htbp]{1.3cm} \centerline{\psfig{figure=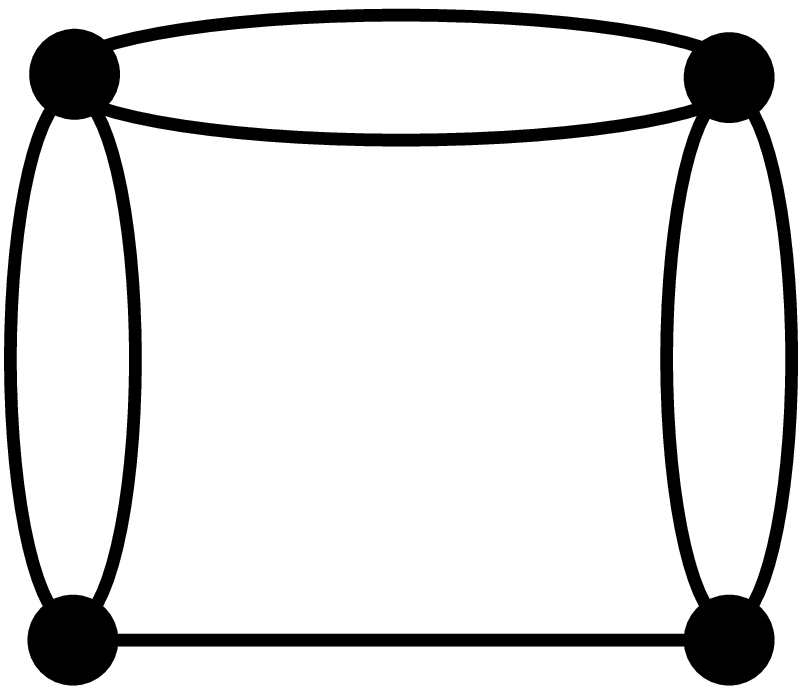,width=1.1cm,angle=0}}\end{minipage} ~ + ~  \ldots \,\, \right]
\right. \nonumber \\
~ &+& ~ 12 \,\, \left[ \,\,   \begin{minipage} [htbp]{1.3cm} \centerline{\psfig{figure=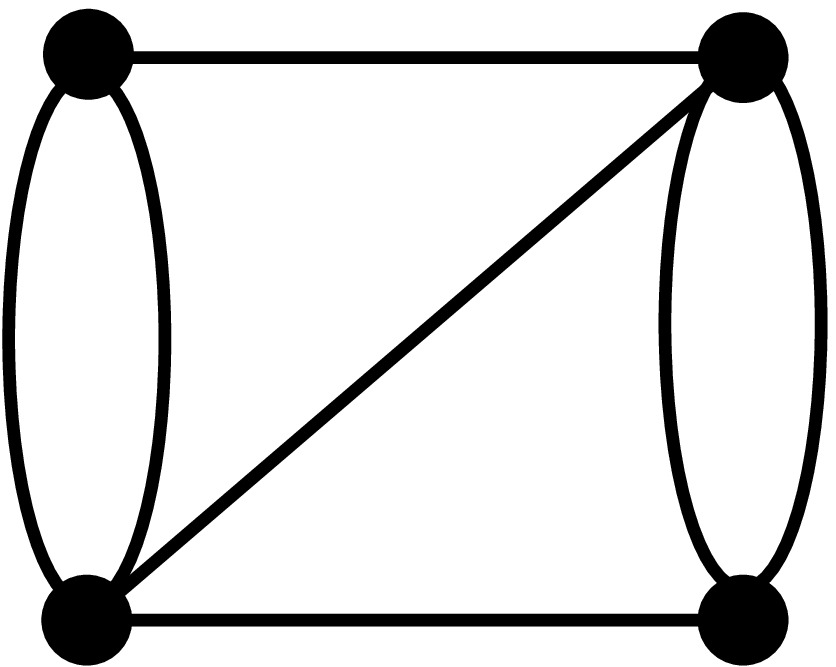,width=1.1cm,angle=0}}\end{minipage} ~ + ~  \ldots \,\, \right] 
~ + ~ 12 \,\, \left[ \,\,   \begin{minipage} [htbp]{1.3cm} \centerline{\psfig{figure=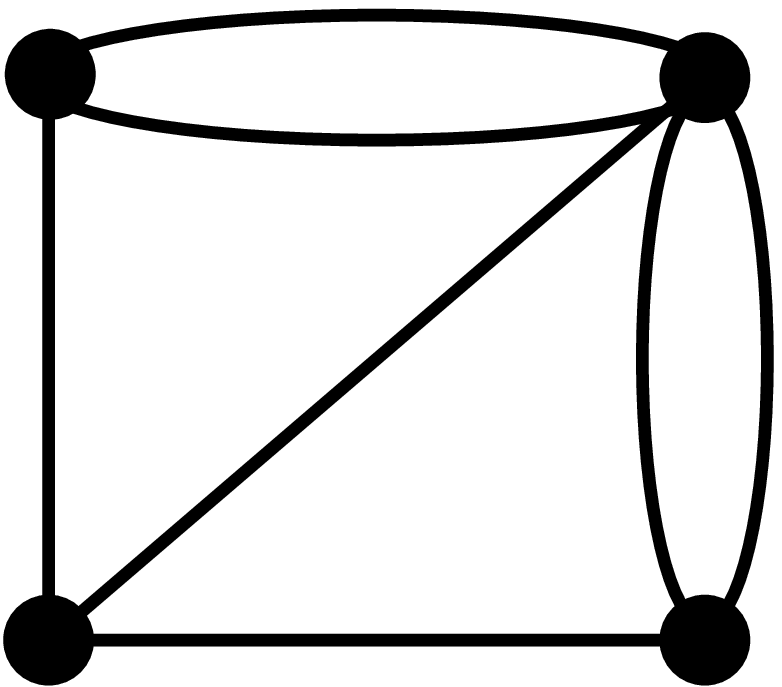,width=1.1cm,angle=0}}\end{minipage} ~ + ~  \ldots \,\, \right]
~ + ~ 3 \,\, \left[ \,\,   \begin{minipage} [htbp]{1.3cm} \centerline{\psfig{figure=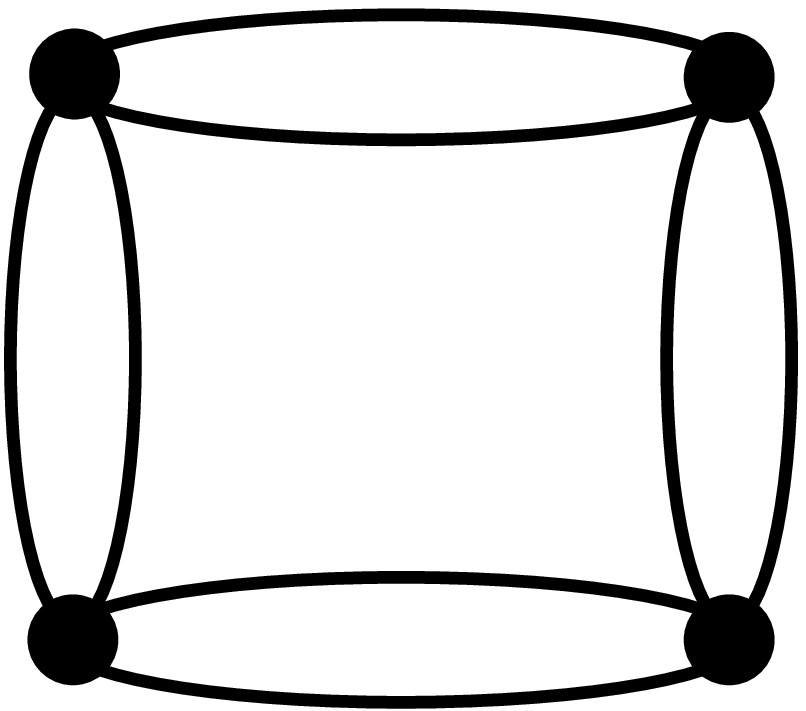,width=1.1cm,angle=0}}\end{minipage} ~ + ~  \ldots \,\, \right]
 \nonumber \\
~ &+& ~ \left. 24 \,\, \left[ \,\,   \begin{minipage} [htbp]{1.3cm} \centerline{\psfig{figure=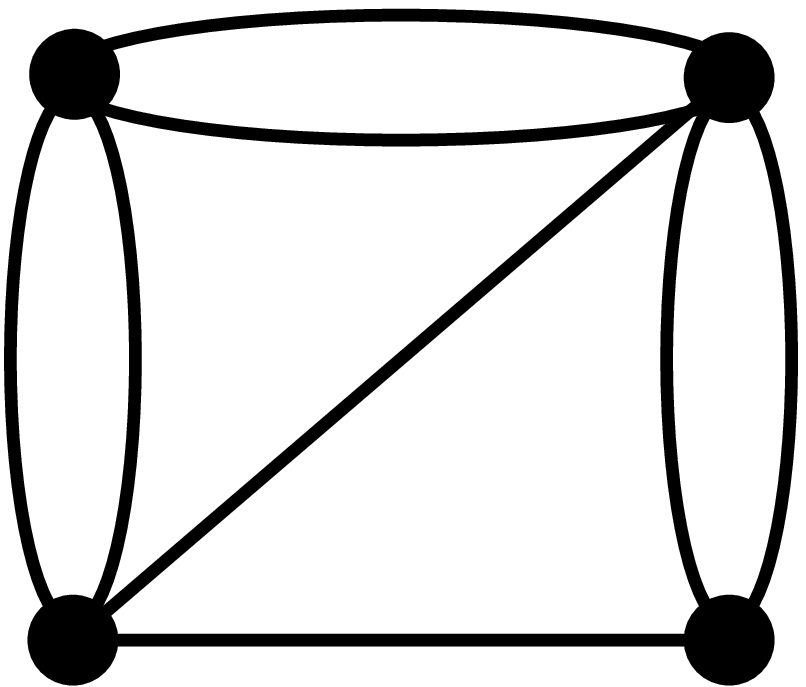,width=1.1cm,angle=0}}\end{minipage} ~ + ~  \ldots \,\, \right]
~ + ~ 6 \,\, \left[ \,\,   \begin{minipage} [htbp]{1.3cm} \centerline{\psfig{figure=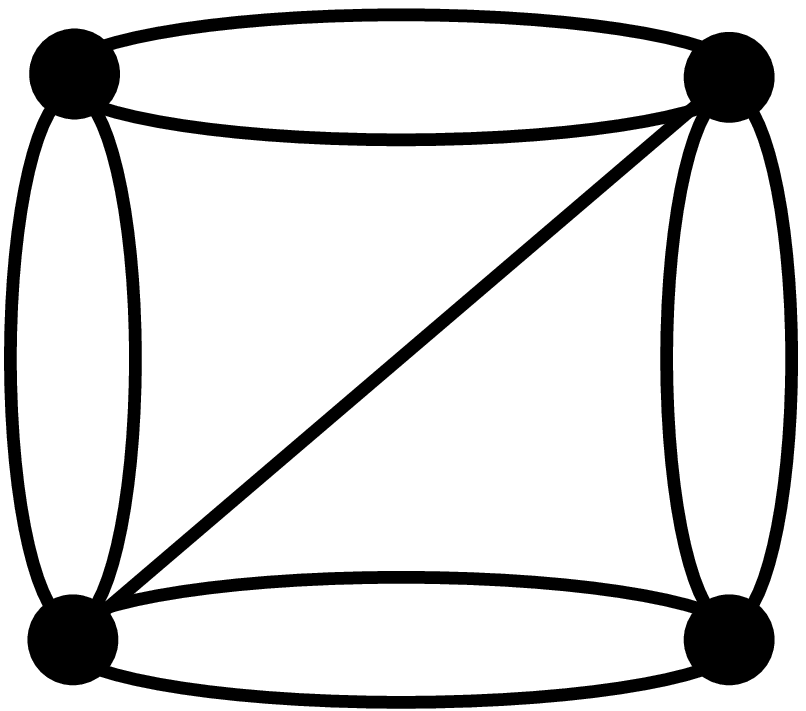,width=1.1cm,angle=0}}\end{minipage} ~ + ~  \ldots \,\, \right]
\, \right\} \, .
\eeq

The dots beyond a diagram of a given type represent all diagrams constructed from the basic one by adding an arbitrary number of solid lines between points connected by 2 lines (for the case of outer lines) or by one line (for the case of inner lines). Introducing a generalized Mayer bond of second type,
$$
\Psi_{ab}(r_{ab})\;=\;\exp[-\,g_{ab}(r_{ab})]\,\,-\,\,1\,\, ,
$$
with the graphic representation,
\be \label{Mayergraph2}
\begin{minipage}[htbp]{1.3cm} \centerline{\psfig{figure=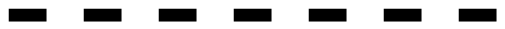,width=1.1cm,angle=0}}\end{minipage}~=~\begin{minipage}[htbp]{1.3cm} \centerline{\psfig{figure=figures/line.eps,width=1.1cm,angle=0}}\end{minipage} ~+~ \begin{minipage}[htbp]{1.3cm} \centerline{\psfig{figure=figures/line2.eps,width=1.1cm,angle=0}}\end{minipage}~+~\begin{minipage}[htbp]{1.3cm} \centerline{\psfig{figure=figures/line3.eps,width=1.1cm,angle=0}}\end{minipage}\,\,  + \,\, \ldots \,\, ~=~ \begin{minipage}[htbp]{1.3cm} \centerline{\psfig{figure=figures/line.eps,width=1.1cm,angle=0}}\end{minipage} ~+~ \begin{minipage}[htbp]{2.3cm} \centerline{\psfig{figure=figures/mayer.eps,width=2.1cm,angle=0}}\end{minipage},
\ee  
one writes the fourth cluster integral in the following manner \cite{F62,Sch68,JG80},
\beq \label{Fourth}
{\cal{S}}_{4}~&=&~ \frac{1}{4!} \,\,\left\{ \,\, 6 \,\,  \begin{minipage} [htbp]{2.3cm} \centerline{\psfig{figure=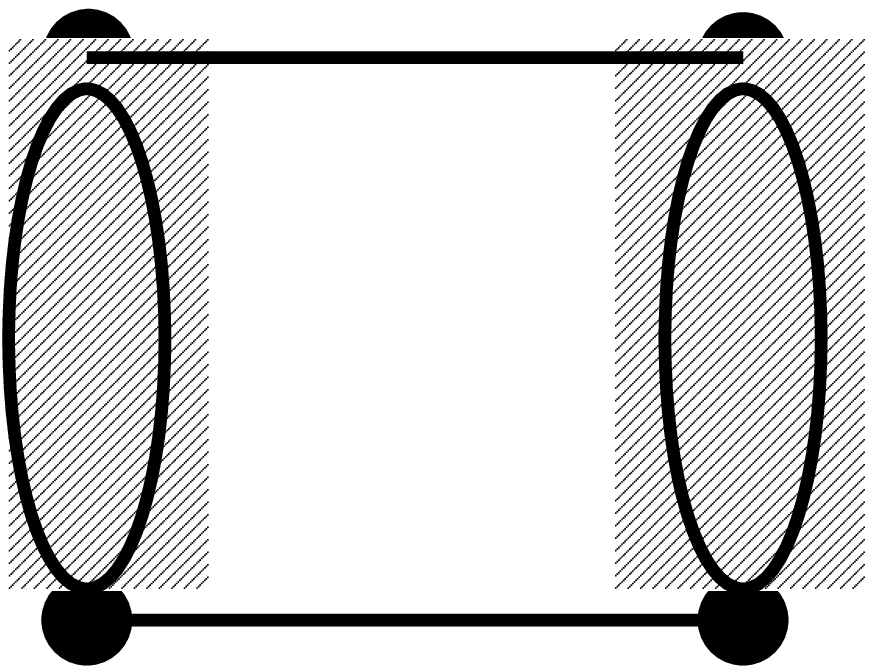,width=2.1cm,angle=0}}\end{minipage} 
~ + ~ \begin{minipage} [htbp]{2.3cm} \centerline{\psfig{figure=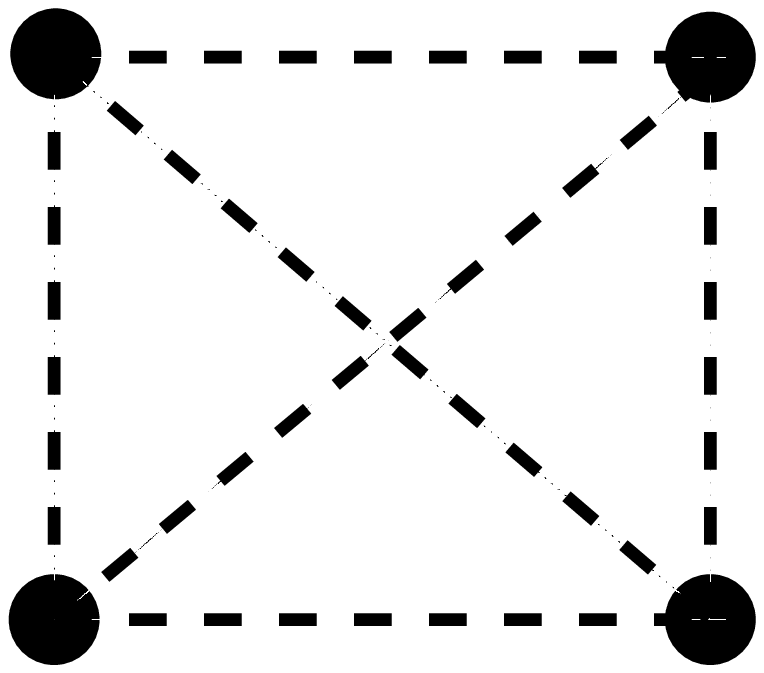,width=2.1cm,angle=0}}\end{minipage} 
~ + ~ 12 \,\,     \begin{minipage} [htbp]{2.3cm} \centerline{\psfig{figure=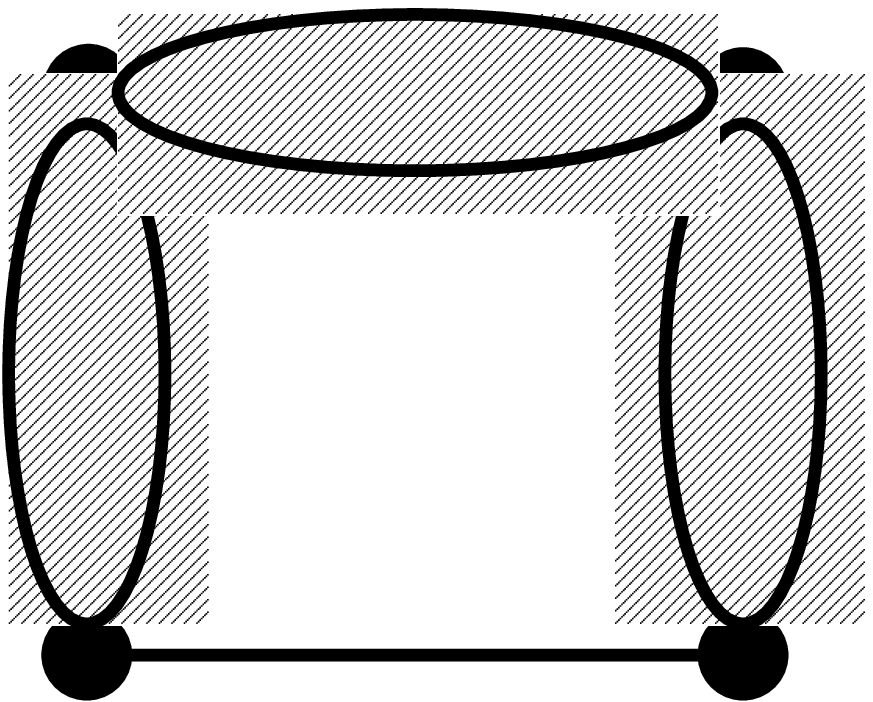,width=2.1cm,angle=0}}\end{minipage}
\right. \nonumber \\ 
~&& + ~ 12 \,\,  \begin{minipage} [htbp]{2.3cm} \centerline{\psfig{figure=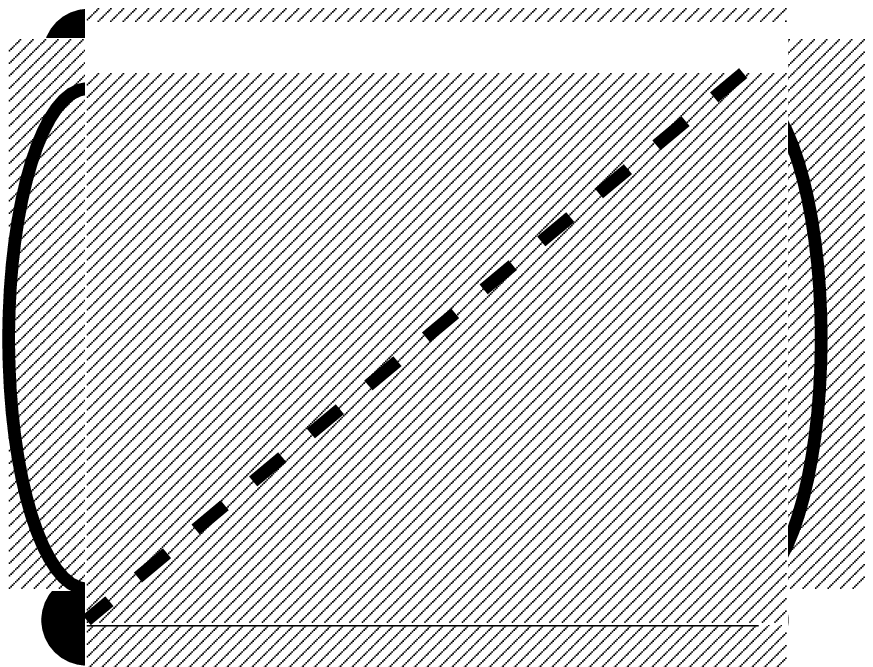,width=2.1cm,angle=0}}\end{minipage}
~+ ~12 \,\,   \begin{minipage} [htbp]{2.3cm} \centerline{\psfig{figure=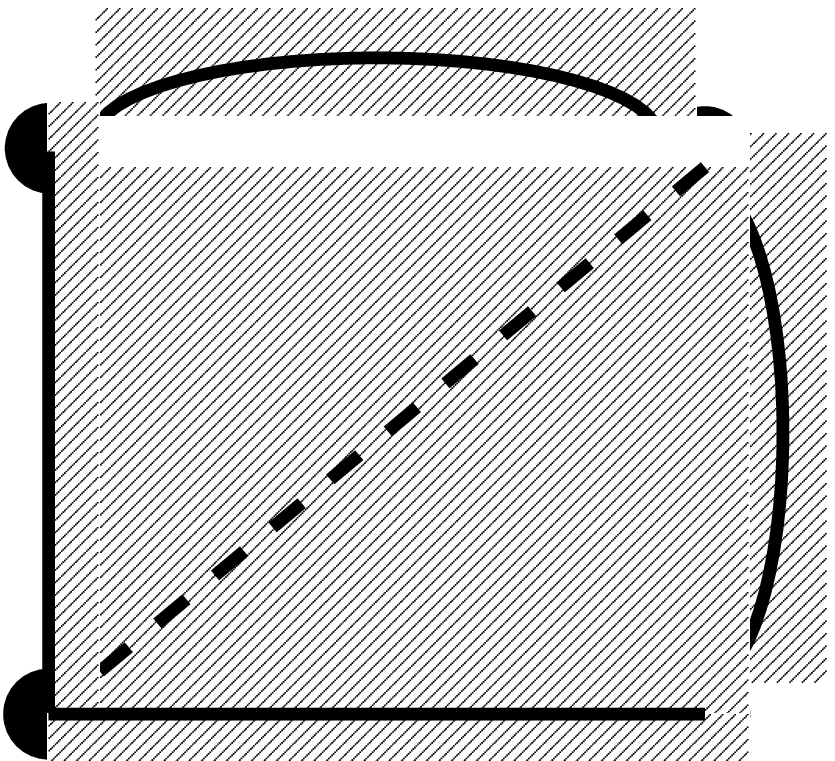,width=2.1cm,angle=0}}\end{minipage}
~ + ~ 3 \,\, \begin{minipage} [htbp]{2.3cm} \centerline{\psfig{figure=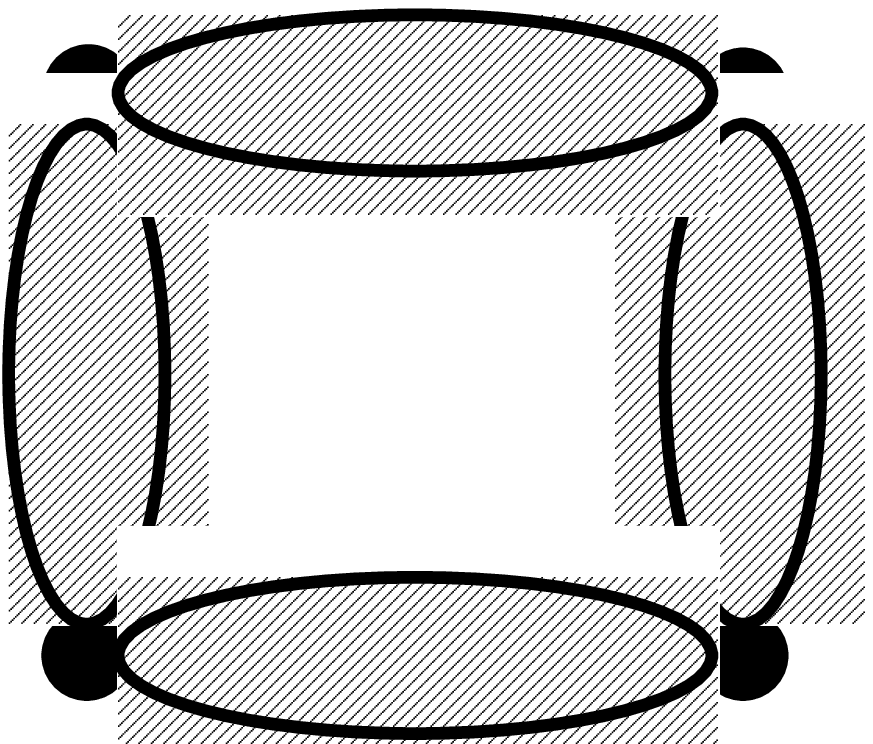,width=2.1cm,angle=0}}\end{minipage}
\nonumber \\ 
~&& +  \left.~  24 \,\,   \begin{minipage} [htbp]{2.3cm} \centerline{\psfig{figure=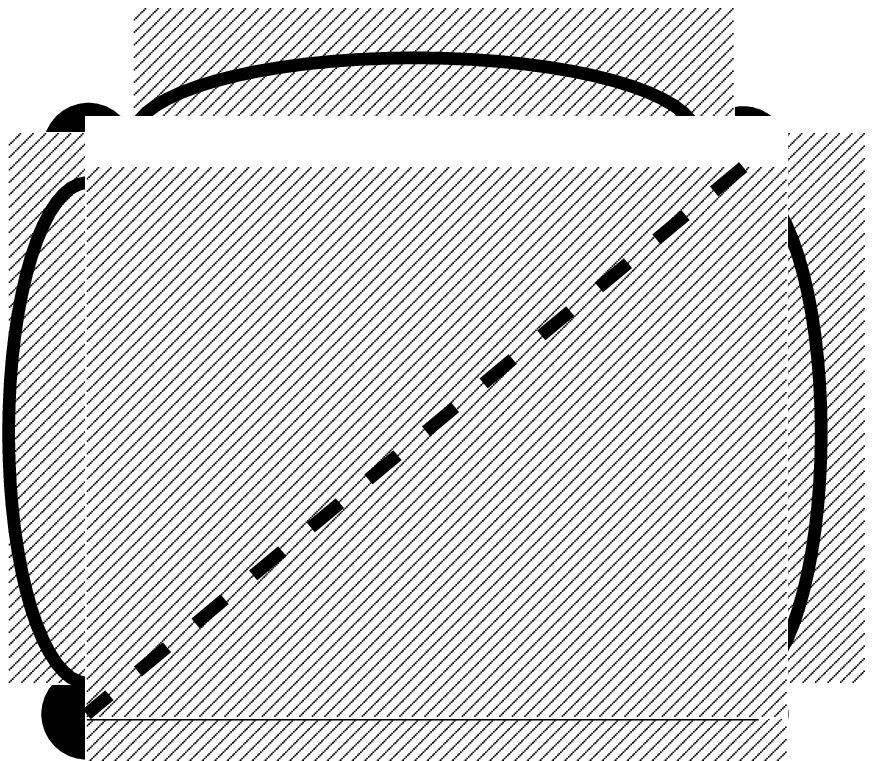,width=2.1cm,angle=0}}\end{minipage}
~ + ~ 6 \,\,   \begin{minipage} [htbp]{2.3cm} \centerline{\psfig{figure=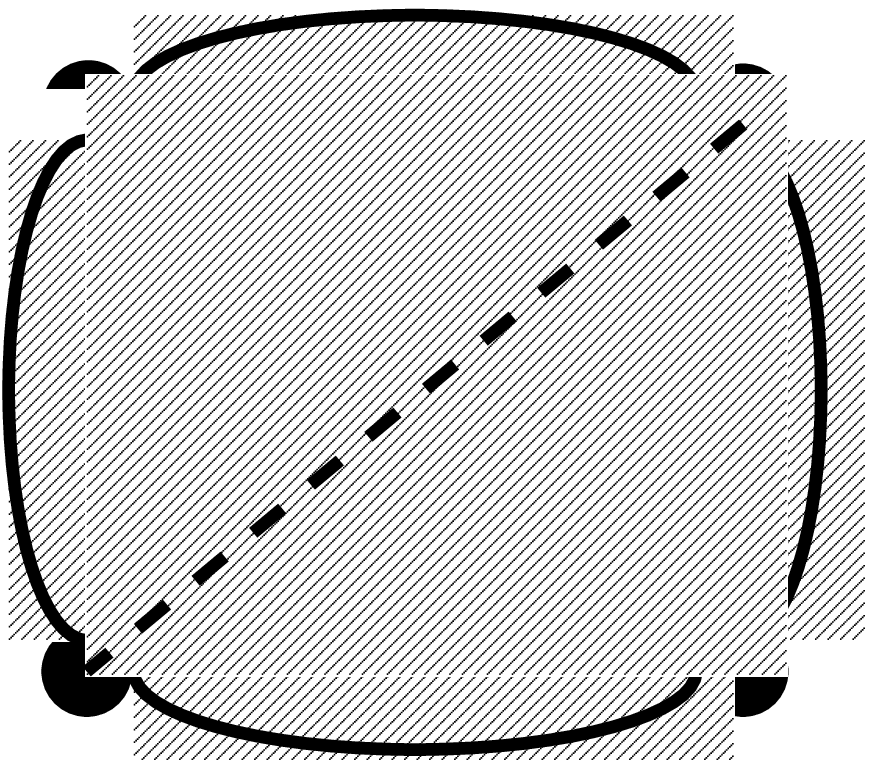,width=2.1cm,angle=0}}\end{minipage}
\, \right\} \, .
\eeq

The leading contribution from the fourth cluster integral are the two graphs with 6 solid lines (stemming from the ${\cal H}_3 \,{\cal H}_3 \,{\cal H}_3 {\cal H}_3$ term), 
\be \label{fourthl}
{\cal{S}}_{4}^l~=~ \frac{1}{4!} \,\,\left\{ \,\, 6 \,\,  \begin{minipage} [htbp]{1.3cm} \centerline{\psfig{figure=figures/fourth1.eps,width=1.1cm,angle=0}}\end{minipage}
~ + ~  \begin{minipage} [htbp]{1.3cm} \centerline{\psfig{figure=figures/fourth2.eps,width=1.1cm,angle=0}}\end{minipage} \right\} \quad .
\ee
According to the rules formulated in the subsection \ref{thirdsection} they are of the order of $n^{5/2}$, and give \cite{Haga,F62}
\be \label{Fourthl}
{\cal{S}}_{4}^l={b_{4}\frac{n^4}{\kappa^{3}}}\beta^6 \,\, \sum_{a,b,c,d}c_{a}c_{b}c_{c}c_{d}
e_a^3e_b^3e_c^3e_d^3 \quad ,
\ee
with the numerical constant $b_{4}=b_{4,\alpha}+b_{4,\beta}$,
\begin{eqnarray}
\label{eq:a2alpha}
b_{4,\alpha}&=&{1\over 16}\int\!\!d{\mathbf{x}}_{1}d{\mathbf{x}}_{2}d{\mathbf{x}}
_{3}\,\biggl({e^{- x_{1}}\over x_{1}}\biggr)^{\!2}\,{e^{-x_{12}}\over x_{12}}
\biggl({e^{-x_{23}}\over x_{23}}\biggr)^{\!2}\,{e^{-x_{3}}\over x_{3}} \quad ,\\[1ex]
\label{eq:a2beta}
b_{4,\beta}&=&{1\over 24}\int\!\!d{\mathbf{x}}_{1}d{\mathbf{x}}_{2}d{\mathbf{x}}
_{3}\,{e^{-x_{1}}\over x_{1}}\,{e^{-x_{2}}\over x_{2}}\,{e^{-x_{3}}\over
x_{3}}\,{e^{-x_{12}}\over x_{12}}\,{e^{-x_{13}}\over x_{13}}\,{e^{-x_{23}}
\over x_{23}}\,.
\end{eqnarray}
The constant $b_{4,\alpha}$ is calculated in Appendix \ref{AppA}. It equals
\be \label{a2alpha}
b_{4 \alpha}=2\pi^{3}\biggl[{\pi^{2}\over 12} + \mathrm{Li}_{2}\biggl(-{1\over 3}\biggr) + \frac{4}{3} \ln{\frac{3}{4}} \biggr]
=8.05281\dots\,.
\ee
The constant $b_{4,\beta}$ was evaluated numerically by a Monte Carlo calculation and equals, in closed correspondence with \cite{Alastuey}
\be \label{a2beta}
b_{4,\beta}\simeq 1.785 \pm 0.001 \,\,.
\ee

Consider now all diagrams with 7 solid lines representing the ${\cal H}_3 \,{\cal H}_3 \,{\cal H}_4 {\cal H}_4$ and ${\cal H}_3 \,{\cal H}_3 \,{\cal H}_3 {\cal H}_5$ terms and contributing to the order $n^3$.
\beq \label{fourthdelta}
\delta {\cal{S}}_{4}~&=&~ \frac{1}{4!} \,\,\left\{ \,\, 12 \,\,  \begin{minipage} [htbp]{1.3cm} \centerline{\psfig{figure=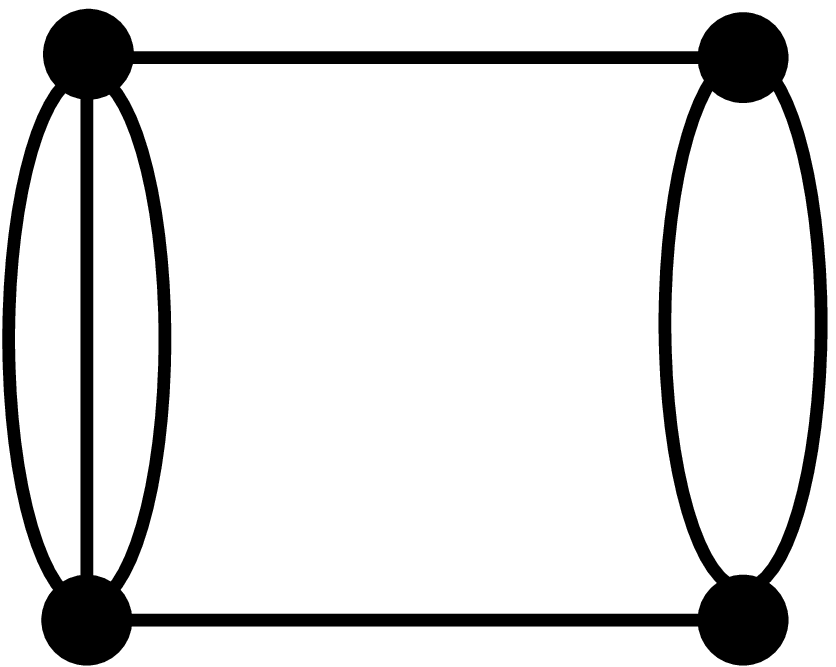,width=1.1cm,angle=0}}\end{minipage}
~ + ~ 6 \,\,  \begin{minipage} [htbp]{1.3cm} \centerline{\psfig{figure=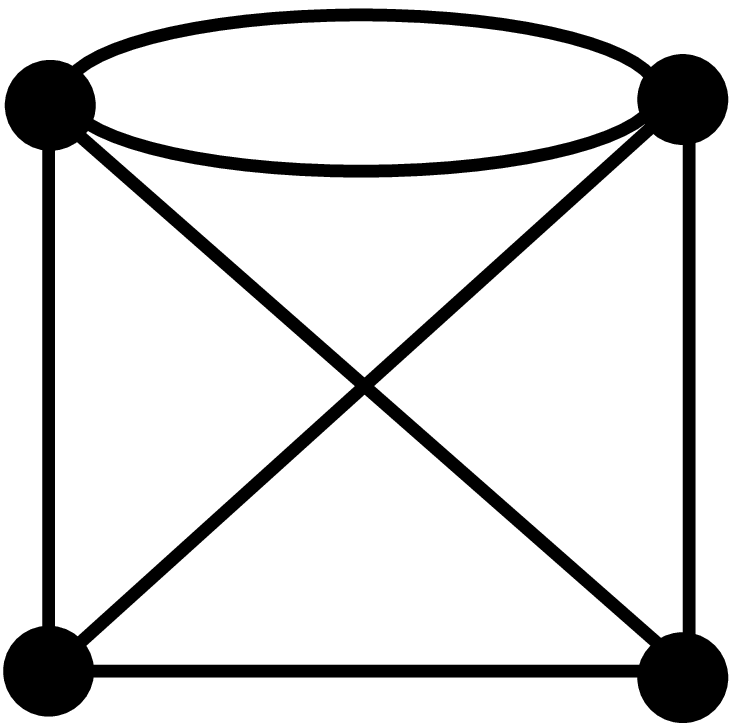,width=1.1cm,angle=0}}\end{minipage}~ + ~ 12 \,\,  \begin{minipage} [htbp]{1.3cm} \centerline{\psfig{figure=figures/fourth3.eps,width=1.1cm,angle=0}}\end{minipage}  
\right. \nonumber \\
&~& + ~ \left. 12 \,\,  \begin{minipage} [htbp]{1.3cm} \centerline{\psfig{figure=figures/fourth4.eps,width=1.1cm,angle=0}}\end{minipage}~ + ~ 12 \,\,  \begin{minipage} [htbp]{1.3cm} \centerline{\psfig{figure=figures/fourth5.eps,width=1.1cm,angle=0}}\end{minipage}~+~o\left(n^3\right) \,\, 
\right\} \quad .
\eeq
Again the first diagram in the above equation is divergent and we may apply the renormalization procedure as described in the section \ref{thirdsection}. The details are given in Appendix \ref{AppC}. We obtain
\beq \label{deltafourth}
\delta {\cal{S}}_{4}~=~\frac{n^4}{\kappa^{2}} \beta^7 \,\, \sum_{a,b,c,d}c_{a}c_{b}c_{c}c_{d}\left\{ e_a^4e_b^4e_c^3e_d^3 \left[\,\, \frac{4 \pi^3}{9}\ln{\kappa\l_{ab}}~-~c_{4}\,\,\right] \right.
~- \left. ~e_a^5e_b^3e_c^3e_d^3 \,\, c_{4\epsilon}\,\,\right\} \,\,,
\eeq
where the constants $c_4~=~c_{4\alpha}+c_{4\beta}+c_{4\gamma}+c_{4\delta}$ and $c_{4\epsilon}$ has been introduced,
\beq \label{eq:c_4alpha}
c_{4\alpha}~&=&~\frac{4 \pi^3}{9}\left[\frac{17}{6}-2C_E\right] ~-~ \frac{16\pi^2}{3} \int_0^{\infty}\frac{dk}{\left(k^2+1\right)^2} \, \arctan{\frac{k}{2}}\,\,\left\{\,\,3 \arctan{\frac{k}{3}} \,\,+\,\,\frac{k}{2}\ln{(k^2+9)}\right\}\,\,,\\[1ex]
c_{4,\beta}&=&{1\over 8}\int\!\!d{\mathbf{x}}_{1}d{\mathbf{x}}_{2}d{\mathbf{x}}
_{3}\,\biggl({e^{- x_{1}}\over x_{1}}\biggr)^{\!2}\,{e^{-x_{2}}\over x_{2}}\,{e^{-x_{3}}\over
x_{3}}\,{e^{-x_{12}}\over x_{12}}\,{e^{-x_{13}}\over x_{13}}\,{e^{-x_{23}}
\over x_{23}}\,
\,,\\[1ex]
c_{4,\gamma}&=&{1\over 16}\int\!\!d{\mathbf{x}}_{1}d{\mathbf{x}}_{2}d{\mathbf{x}}
_{3}\,\biggl({e^{- x_{1}}\over x_{1}}\biggr)^{\!2}\,\biggl({e^{- x_{3}}\over x_{3}}\biggr)^{\!2}\,\biggl({e^{- x_{12}}\over x_{12}}\biggr)^{\!2}\,{e^{-x_{23}}
\over x_{23}}\,\,,
\\[1ex]
c_{4,\delta}&=&{1\over 8}\int\!\!d{\mathbf{x}}_{1}d{\mathbf{x}}_{2}d{\mathbf{x}}
_{3}\,{e^{- x_{1}}\over x_{1}}\,\biggl({e^{- x_{3}}\over x_{3}}\biggr)^{\!2}\,\biggl({e^{- x_{12}}\over x_{12}}\biggr)^{\!2}\,{e^{-x_{23}}
\over x_{23}}\, {e^{-x_{13}} \over x_{13}}
\,\,,\\[1ex]
c_{4,\epsilon}&=&{1\over 8}\int\!\!d{\mathbf{x}}_{1}d{\mathbf{x}}_{2}d{\mathbf{x}}
_{3}\,\biggl({e^{- x_{1}}\over x_{1}}\biggr)^{\!2}\,\biggl({e^{- x_{3}}\over x_{3}}\biggr)^{\!2}\,{e^{- x_{12}}\over x_{12}}\,{e^{-x_{23}}
\over x_{23}}\,{e^{-x_{2}} \over x_{2}}\,\,.
\,\, 
\eeq
The above multiple integrals for $c_{4,\beta}$, $c_{4,\delta}$ and $c_{4,\epsilon}$ can be reduced to single or double integrals and are evaluated by numerical integration. The remaining integral is estimated by a Monte Carlo run.  As a result we obtain:
\beq \label{c4number}
c_4~&=&~ 53.66 \pm 0.02\,,
\nonumber \\
c_{4,\epsilon}&=& 26.719 \dots\,\,.
\eeq
The fifth and sixth cluster integral contributing to the order $n^3$ are considered in the next section.
\subsection{Fifth and sixth cluster integral (leading terms)}
The diagrammatic representation of the fifth and sixth cluster integral is given by a huge number of basic diagrams. Thus the fifth cluster integral is represented in Ref. \cite{JG80}. According to our aim of presenting a virial expansion of the free energy it suffices to give the leading contributions to both cluster integrals.

The leading contribution of the fifth cluster integral is given by the four graphs with 8 solid lines (stemming from the ${\cal H}_4 \,\left({\cal H}_3 \right)^4$ term), 
\be \label{fifthl}
{\cal{S}}_{5}^l~=~ \frac{1}{5!} \,\,\left\{ \,\, 15 \,\,  \begin{minipage} [htbp]{1.3cm} \centerline{\psfig{figure=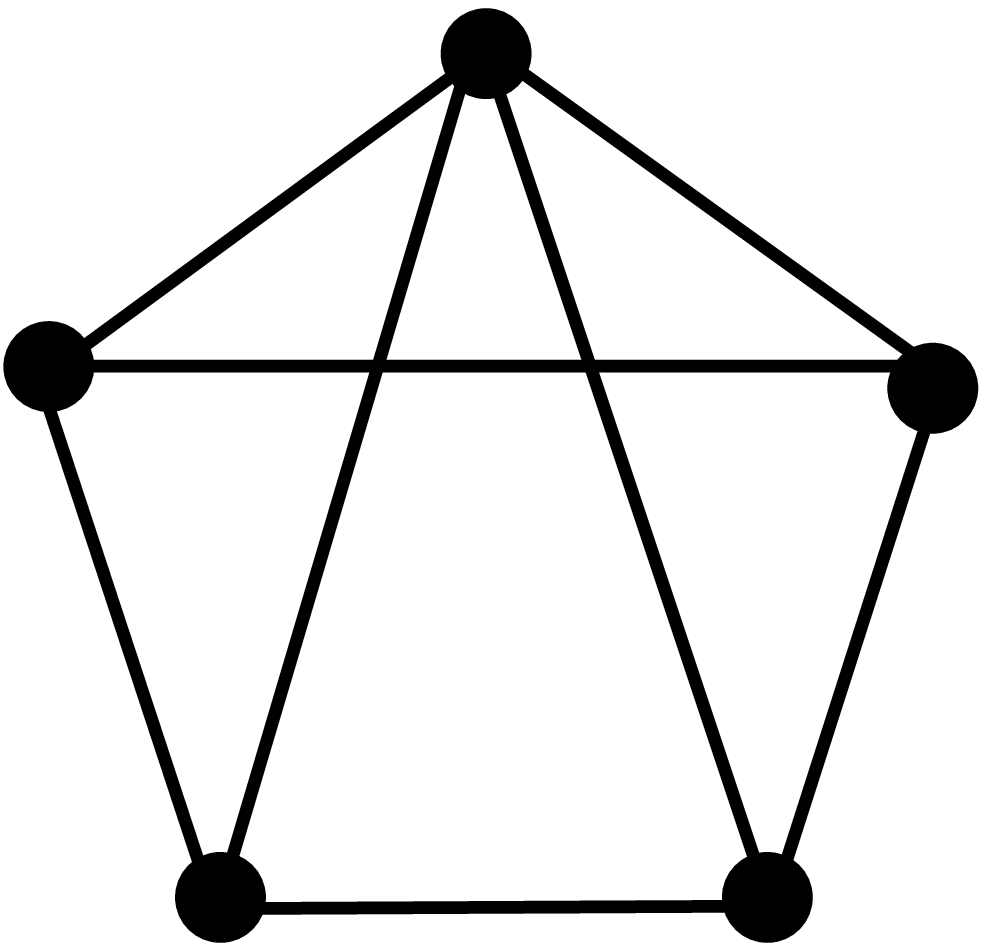,width=1.1cm,angle=0}}\end{minipage}
~ + ~  60 \,\, \begin{minipage} [htbp]{1.3cm} \centerline{\psfig{figure=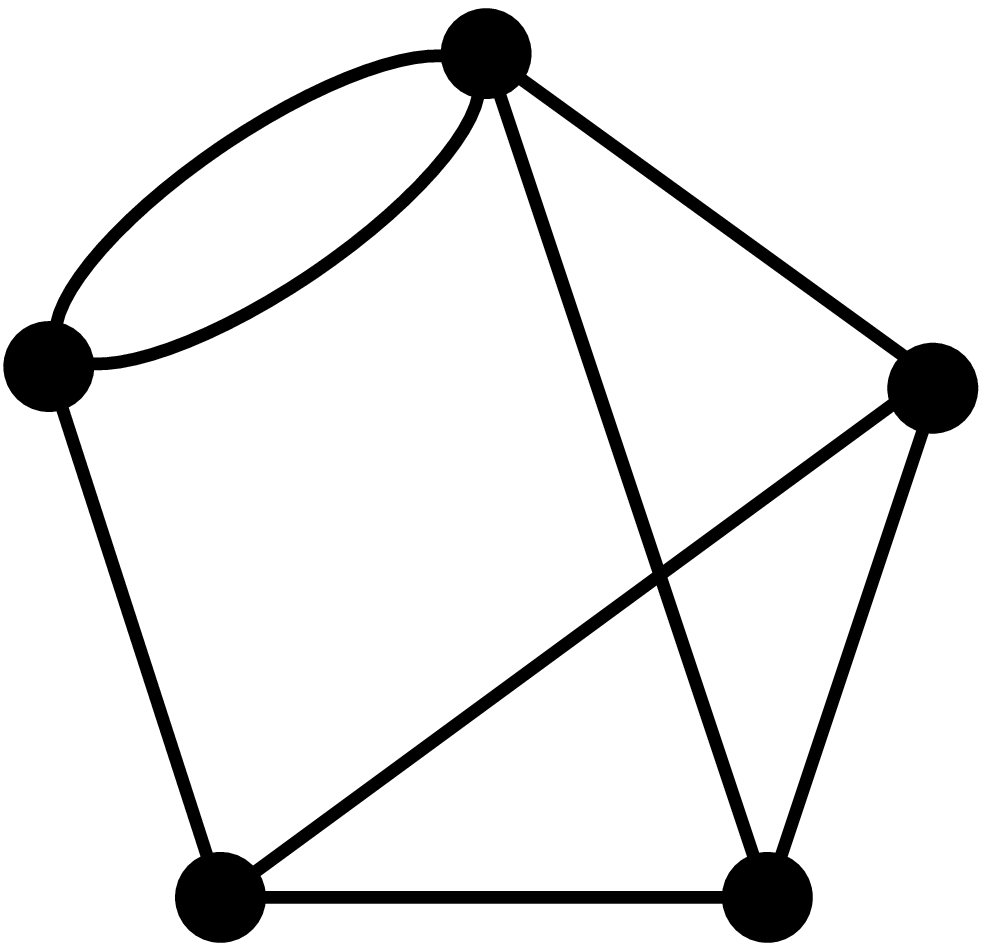,width=1.1cm,angle=0}}\end{minipage}~ + ~  60 \,\, \begin{minipage} [htbp]{1.3cm} \centerline{\psfig{figure=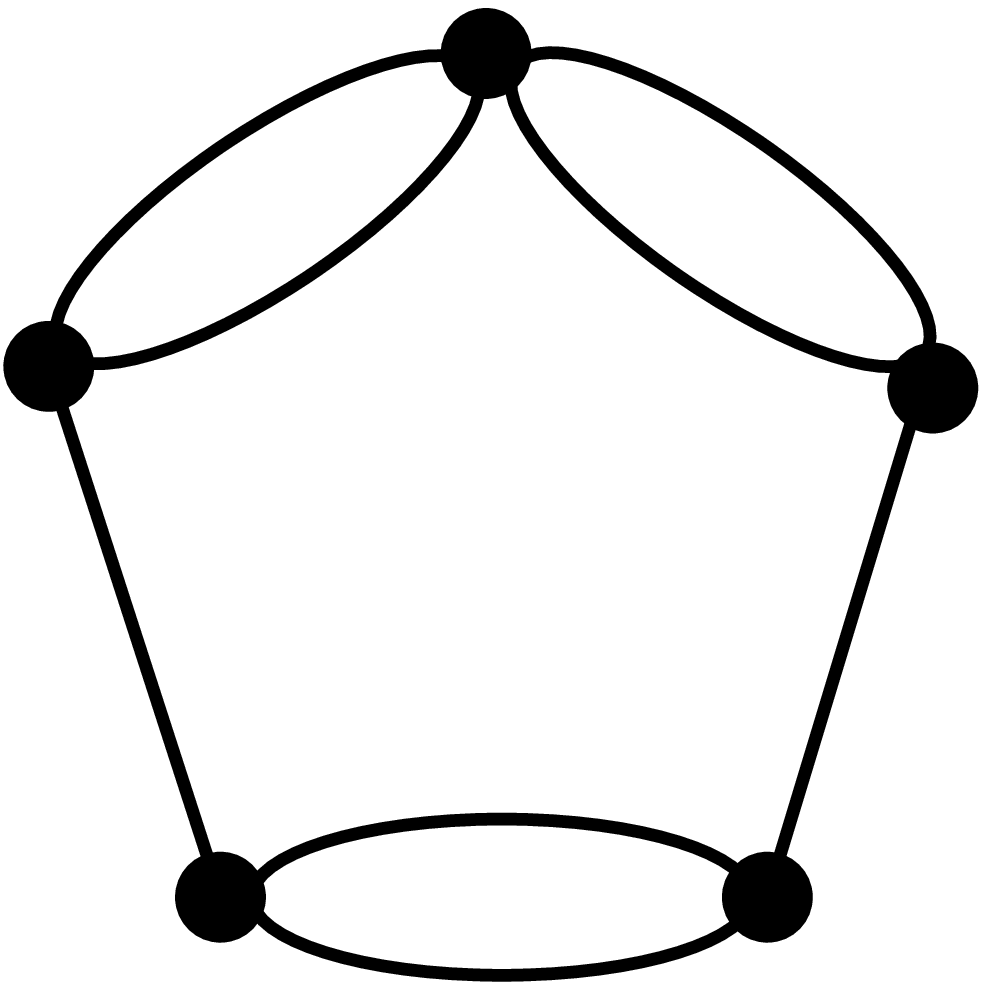,width=1.1cm,angle=0}}\end{minipage} ~ + ~  120 \,\, \begin{minipage} [htbp]{1.3cm} \centerline{\psfig{figure=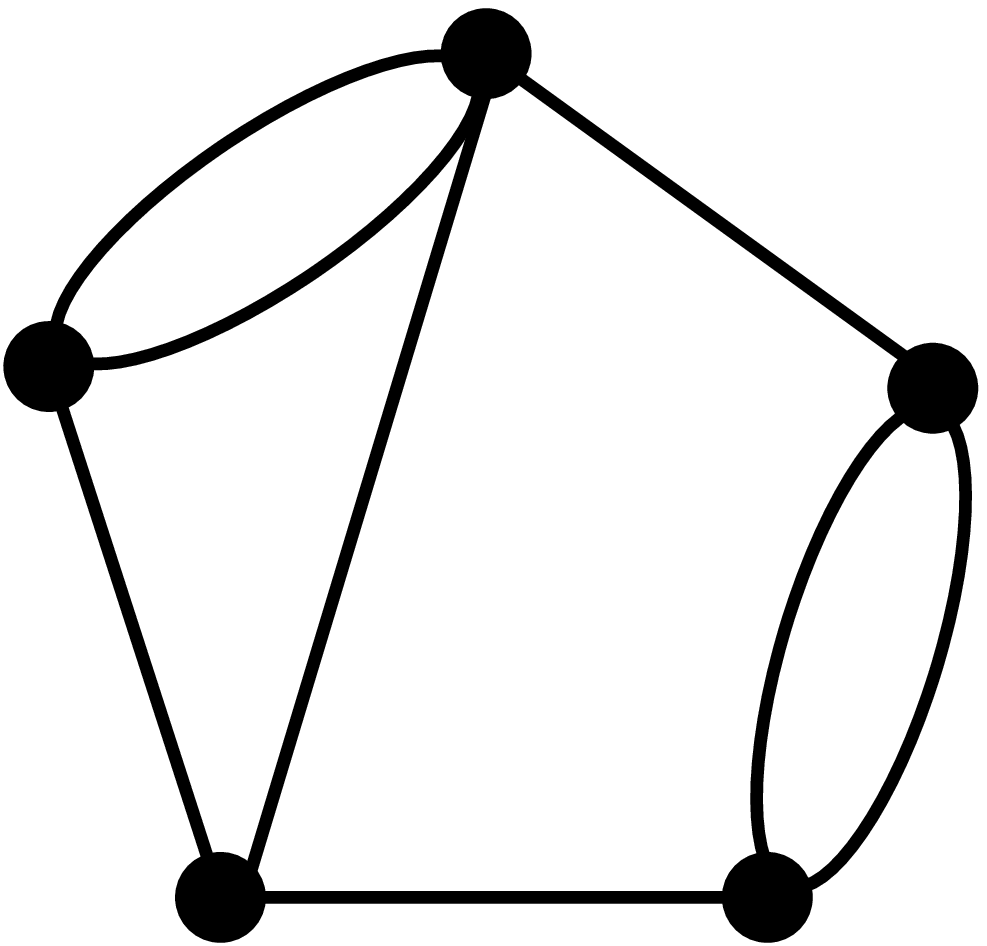,width=1.1cm,angle=0}}\end{minipage}\right\} \quad .
\ee
According to the rules formulated in the subsection \ref{thirdsection} they are of the order of $n^{3}$, and give 
\be \label{Fifthl}
{\cal{S}}_{5}^l=\,{c_{5}\frac{n^5}{\kappa^{4}}} \beta^8 \,\, \sum_{a,b,c,d,f}c_{a}c_{b}c_{c}c_{d}c_{f}
e_a^4e_b^3e_c^3e_d^3e_f^3 \quad ,
\ee
with the numerical constant $c_{5}=c_{5,\alpha}+c_{5,\beta}+c_{5,\gamma}+c_{5,\delta}$.

The four constants introduced here are determined by the dimensionless integrals.
\begin{eqnarray}
\label{eq:b1alpha}
c_{5,\alpha}&=&{1\over 8}\int\!\!d{\mathbf{x}}_{1}d{\mathbf{x}}_{2}d{\mathbf{x}}
_{3}d{\mathbf{x}}_{4}\,{e^{-x_{1}}\over x_{1}}\,{e^{-x_{12}}\over x_{12}}\,{e^{-x_{2}}\over x_{2}}\,{e^{-x_{23}}\over x_{23}}\,{e^{-x_{3}}\over x_{3}}\,{e^{-x_{34}}\over x_{34}}\,{e^{-x_{4}}\over x_{4}}\,{e^{-x_{41}}\over x_{41}} \quad,\\[1ex]
\label{eq:b1beta}
c_{5,\beta}&=&{1\over 4}\int\!\!d{\mathbf{x}}_{1}d{\mathbf{x}}_{2}d{\mathbf{x}}
_{3}d{\mathbf{x}}_{4}\,\biggl({e^{-x_{1}}\over x_{1}}\biggr)^{\!2}\,{e^{-x_{2}}\over x_{2}}\,{e^{-x_{4}}\over x_{4}}\,{e^{-x_{13}}\over x_{13}}\,{e^{-x_{34}}\over x_{34}}\,{e^{-x_{42}}\over x_{42}}\,{e^{-x_{23}}\over x_{23}}\,
\,,\\[1ex]
\label{eq:b1gamma}
c_{5,\gamma}&=&{1\over 16}\int\!\!d{\mathbf{x}}_{1}d{\mathbf{x}}_{2}d{\mathbf{x}}
_{3}d{\mathbf{x}}_{4}\,\biggl({e^{-x_{1}}\over x_{1}}\biggr)^{\!2}\,\biggl({e^{-x_{2}}\over x_{2}}\biggr)^{\!2}\,{e^{-x_{13}}\over x_{13}}\,{e^{-x_{3}}\over x_{3}}\,{e^{-x_{34}}\over x_{34}}\,
\biggl({e^{-x_{24}}\over x_{24}}\biggr)^{\!2}\,\,,
\\[1ex]
\label{eq:b1delta}
c_{5,\delta}&=&{1\over 4}\int\!\!d{\mathbf{x}}_{1}d{\mathbf{x}}_{2}d{\mathbf{x}
}_{3}d{\mathbf{x}}_{4}\,\biggl({e^{-x_{1}}\over x_{1}}\biggr)^{\!2}\,{e^{-x_{2}}\over x_{2}}\,{e^{-x_{13}}\over x_{13}}\,{e^{-x_{3}}\over x_{3}}\,{e^{-x_{34}}\over x_{34}}\,\biggl({e^{-x_{24}}\over x_{24}}\biggr)^{\!2}\,
\,\, .
\end{eqnarray}
A numerical evaluation (the ``irreducible'' integrals were estimated by Monte Carlo simulations) gives:
\beq \label{c5number}
c_5~&=&~ 168.2 \pm 0.6\,\,.
\eeq

In the sixth cluster integral we can pick out the contributions stemming from the  $\left({\cal H}_3 \right)^6$ term, 
\be \label{sixthl}
{\cal{S}}_{6}^l~=~ \frac{1}{6!} \,\,\left\{ \,\, 120 \,\,  \begin{minipage} [htbp]{1.3cm} \centerline{\psfig{figure=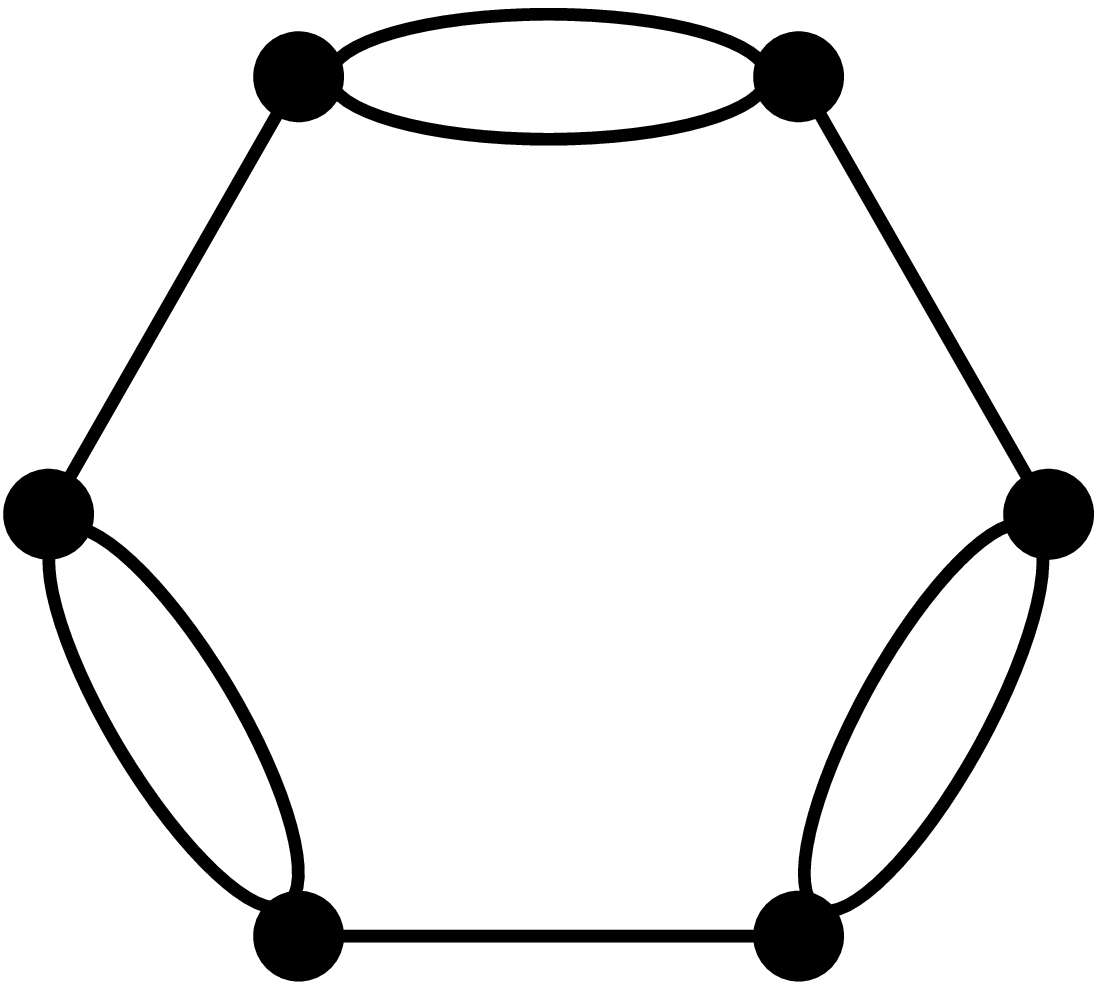,width=1.1cm,angle=0}}\end{minipage}
~ + ~  180 \,\, \begin{minipage} [htbp]{1.3cm} \centerline{\psfig{figure=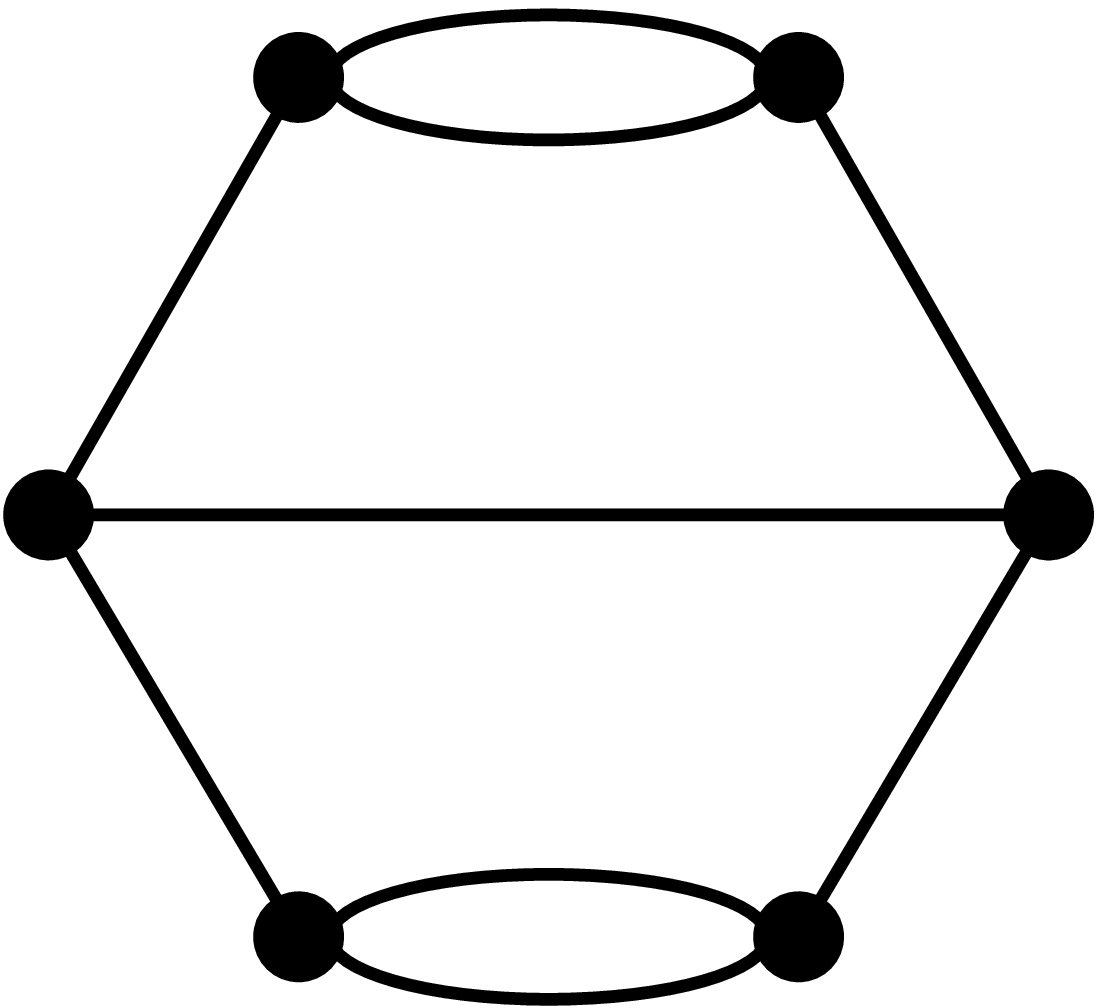,width=1.1cm,angle=0}}\end{minipage}~ + ~  180 \,\, \begin{minipage} [htbp]{1.3cm} \centerline{\psfig{figure=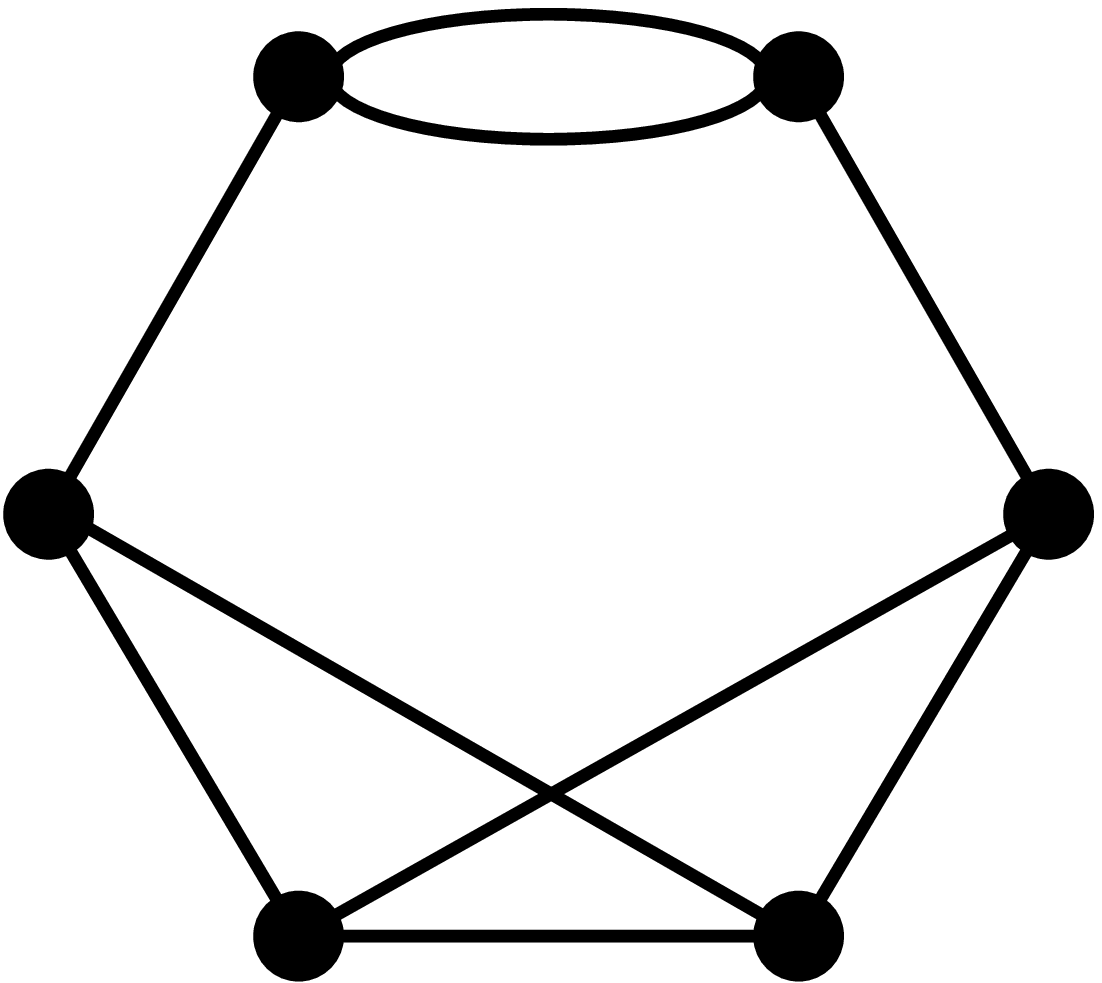,width=1.1cm,angle=0}}\end{minipage} ~ + ~  10 \,\, \begin{minipage} [htbp]{1.3cm} \centerline{\psfig{figure=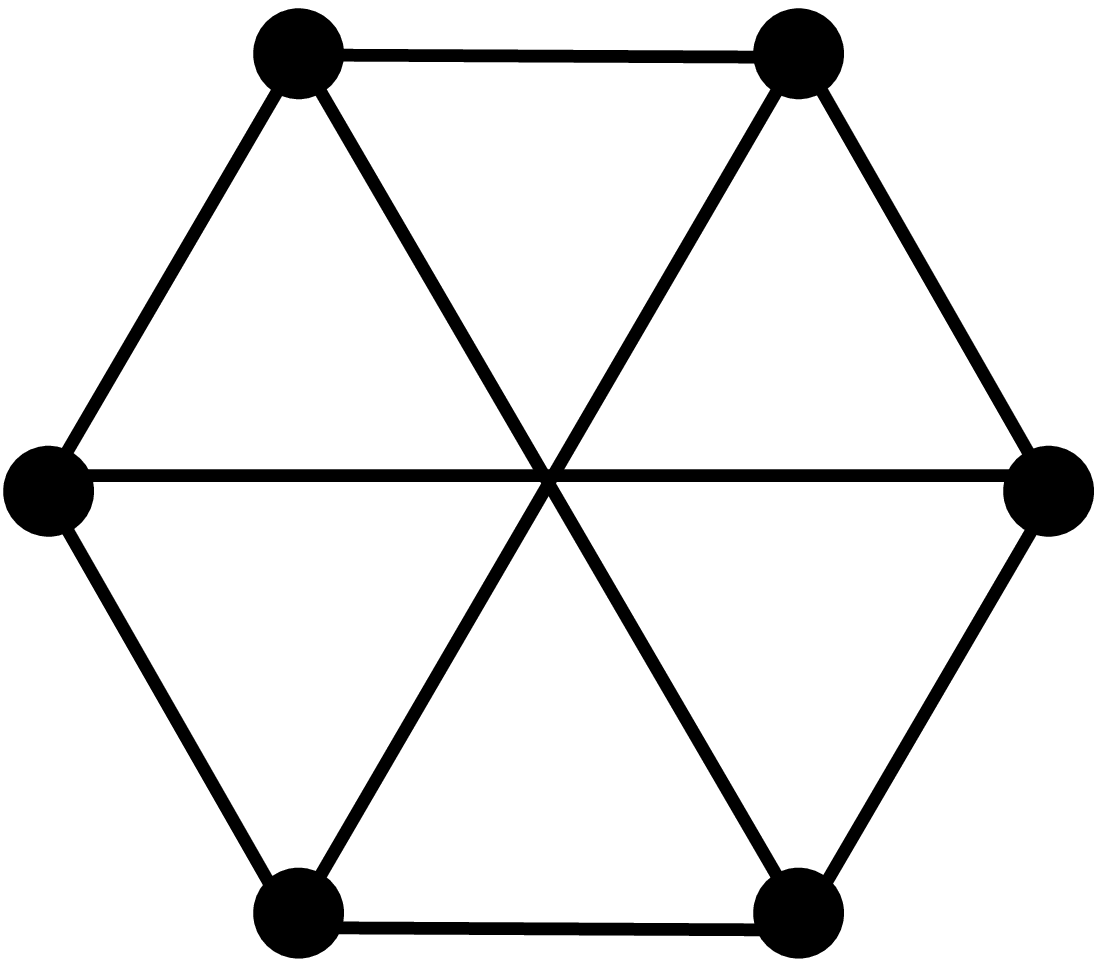,width=1.1cm,angle=0}}\end{minipage}~ + ~  180 \,\, \begin{minipage} [htbp]{1.3cm} \centerline{\psfig{figure=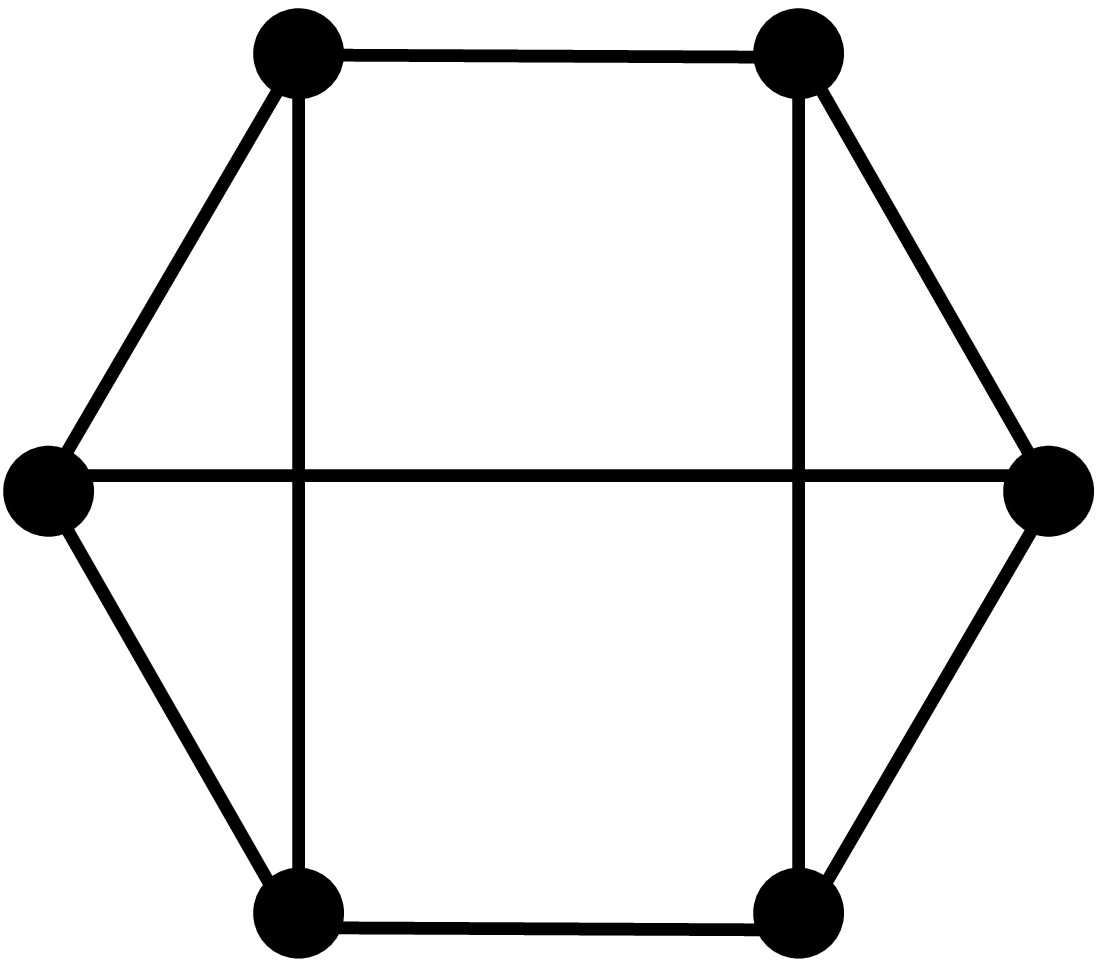,width=1.1cm,angle=0}}\end{minipage}\right\}\quad .
\ee
The five contributions are again of the order $n^{3}$, and give 
\be \label{Sixthl}
{\cal{S}}_{6}^l=~-~{c_{6}\frac{n^6}{\kappa^{6}}} \beta^9 \,\, \sum_{a,b,c,d,f,g}c_{a}c_{b}c_{c}c_{d}c_{f}c_{g}
e_a^3e_b^3e_c^3e_d^3e_f^3e_g^3 \quad,
\ee
with the numerical constant $c_{6}=c_{6,\alpha}+c_{6,\beta}+c_{6,\gamma}+c_{6,\delta}+c_{6,\epsilon}$ given by the five dimensionless integrals,
\begin{eqnarray}
\label{eq:b2alpha}
c_{6,\alpha}&=&{1\over 48}\int\!\!d{\mathbf{x}}_{1}d{\mathbf{x}}_{2}d{\mathbf{x}}
_{3}d{\mathbf{x}}_{4}d{\mathbf{x}}_{5}\,\biggl({e^{-x_{1}}\over x_{1}}\biggr)^{\!2}\,{e^{-x_{12}}\over x_{12}}\,\biggl({e^{-x_{23}}\over x_{23}}\biggr)^{\!2}\,{e^{-x_{34}}\over x_{34}}\,\biggl({e^{-x_{45}}\over x_{45}}\biggr)^{\!2}\,{e^{-x_{5}}\over x_{5}}\,\,,\\[1ex]
\label{eq:b2beta}
c_{6,\beta}&=&{1\over 16}\int\!\!d{\mathbf{x}}_{1}d{\mathbf{x}}_{2}d{\mathbf{x}
}_{3}d{\mathbf{x}}_{4}d{\mathbf{x}}_{5}\,\biggl({e^{-x_{1}}\over x_{1}}\biggr)^{\!2}\,{e^{-x_{12}}\over x_{12}}\,{e^{-x_{23}}\over x_{23}}\,\biggl({e^{-x_{34}}\over x_{34}}\biggr)^{\!2}\,{e^{-x_{45}}\over x_{45}}\,{e^{-x_{5}}\over x_{5}}\,{e^{-x_{52}}\over x_{52}}
\,\,,\\[1ex]
\label{eq:b2gamma}
c_{6,\gamma}&=&{1\over 8}\int\!\!d{\mathbf{x}}_{1}d{\mathbf{x}}_{2}d{\mathbf{x}
}_{3}d{\mathbf{x}}_{4}d{\mathbf{x}}_{5}\,\biggl({e^{-x_{1}}\over x_{1}}\biggr)^{\!2}\,{e^{-x_{12}}\over x_{12}}\,{e^{-x_{23}}\over x_{23}}\,{e^{-x_{34}}\over x_{34}}\,{e^{-x_{45}}\over x_{45}}\,{e^{-x_{5}}\over x_{5}}\,{e^{-x_{53}}\over x_{53}}\,{e^{-x_{24}}\over x_{24}}
\\[1ex]
\label{eq:b2delta}
c_{6,\delta}&=&{1\over 72}\int\!\!d{\mathbf{x}}_{1}d{\mathbf{x}}_{2}d{\mathbf{x}}
_{3}d{\mathbf{x}}_{4}d{\mathbf{x}}_{5}\,{e^{-x_{1}}\over x_{1}}\,{e^{-x_{12}}\over x_{12}}\,{e^{-x_{23}}\over x_{23}}\,{e^{-x_{34}}\over x_{34}}\,{e^{-x_{45}}\over x_{45}}\,{e^{-x_{5}}\over x_{5}}\,{e^{-x_{52}}\over x_{52}}\,{e^{-x_{14}}\over x_{14}}\,{e^{-x_{3}}\over x_{3}} \,\,,
\\[1ex]
\label{eq:b2epsilon}
c_{6,\epsilon}&=&{1\over 4}\int\!\!d{\mathbf{x}}_{1}d{\mathbf{x}}_{2}d{\mathbf{x}}
_{3}d{\mathbf{x}}_{4}d{\mathbf{x}}_{5}\,{e^{-x_{1}}\over x_{1}}\,{e^{-x_{12}}\over x_{12}}\,{e^{-x_{23}}\over x_{23}}\,{e^{-x_{34}}\over x_{34}}\,{e^{-x_{45}}\over x_{45}}\,{e^{-x_{5}}\over x_{5}}\,{e^{-x_{51}}\over x_{51}}\,{e^{-x_{24}}\over x_{24}}\,{e^{-x_{3}}\over x_{3}}
\,\, .
\end{eqnarray}
The numeric evaluation results in:
\beq \label{c6number}
c_6~&=&~ 339.8 \pm 0.3\,.
\eeq

According to the rules formulated in Sec. \ref{thirdsection} there are no other terms contributing to the order $n^3$.  
Thus in this section we have explicitely calculated all terms of the free energy of a classical plasma consisting of charges with the same sign and immersed in a neutralizing background up to the third order in density. The final result is given in Sec. \ref{last}. 

\section{Final results and conclusions} \label{last}

The final expression for the interaction part of the Helmholtz free-energy density to the order $n^3$ is obtained from Eq.(\ref{F_c}) by summing the corresponding contributions ${\cal{S}}_{2}$ (Eq.(\ref{second-final})), ${\cal{S}}_{3}$ (Eqs.(\ref{S3l}),(\ref{a_1}),(\ref{delta1a}),(\ref{c_3alpha}),(\ref{delta2b}),(\ref{c_3beta})), ${\cal{S}}_{4}$ (Eqs.(\ref{fourthl}),(\ref{a2alpha}),(\ref{a2beta}),(\ref{deltafourth}),(\ref{c4number})), ${\cal{S}}_{5}$ (Eqs.(\ref{Fifthl}),(\ref{c5number})), and ${\cal{S}}_{6}$ (Eqs.(\ref{Sixthl}),(\ref{c6number})) with the following result
\beq \label{final}
{\beta F_{\mathrm{c}}\over V}=&-& {\kappa^{3}\over 12\pi}~-~n^2 \beta^3 \sum_{a,b}c_ac_be_a^3e_b^3\left[\frac{\pi}{3}\ln{\kappa\l_{ab}}\,\,+\,\,a_2\right]~-~n^2 \kappa \beta^4 \sum_{a,b}c_ac_be_a^4e_b^4\left[\frac{\pi}{3}\ln{\kappa\l_{ab}}\,\,+\,\,b_2\right]
\nonumber\\
&+&\,\,  b_3 \, n^3 \kappa^{-1} \beta^5 \sum_{a,b,c}c_ac_bc_ce_a^3e_b^4e_c^3~-~ b_4 \, n^4 \kappa^{-3} \beta^6 \sum_{a,b,c,d}c_ac_bc_cc_de_a^3e_b^3e_c^3e_d^3
\nonumber\\
&-& n^3  \beta^6 \sum_{a,b,c}c_ac_bc_ce_a^4e_b^5e_c^3\left[\,\, \frac{2 \pi^2}{3} \ln^2{\kappa l_{ab}} + \frac{8 \pi^2}{3} \left( \, C_E - \frac{17}{12}+ \frac{1}{2}\ln{3}\,\right) \ln{\kappa l_{ab}} ~+~ c_{3\beta} \,\right]
\nonumber\\
&-&~ n^2 \kappa^2 \beta^5 \sum_{a,b}c_ac_be_a^5e_b^5\left[\frac{5\pi}{24}\ln{\kappa\l_{ab}}\,\,+\,\,c_2\right] ~+~  n^3 \beta^6 \sum_{a,b,c}c_ac_bc_ce_a^4e_b^4e_c^4 \left[\frac{\pi^4}{12}\ln{\kappa\l_{ab}}\,\,+\,\,c_{3,\alpha}\right]
\nonumber\\
&-&n^4 \kappa^{-2} \beta^7 \,\, \sum_{a,b,c,d}c_{a}c_{b}c_{c}c_{d}\left\{ e_a^4e_b^4e_c^3e_d^3 \left[\,\, \frac{4 \pi^3}{9}\ln{\kappa\l_{ab}}~-~c_{4}\,\,\right] \right.~- \left. ~e_a^5e_b^3e_c^3e_d^3 \,\, c_{4\epsilon}\,\,\right\} 
\nonumber\\
&-&{c_{5}\frac{n^5}{\kappa^{4}}} \beta^8 \,\, \sum_{a,b,c,d,f}c_{a}c_{b}c_{c}c_{d}c_{f}e_a^4e_b^3e_c^3e_d^3e_f^3~+~{c_{6}\frac{n^6}{\kappa^{6}}} \beta^9 \,\, \sum_{a,b,c,d,f,g}c_{a}c_{b}c_{c}c_{d}c_{f}c_{g}e_a^3e_b^3e_c^3e_d^3e_f^3e_g^3 ~+~o\left(n^3\right)\quad ,
\eeq
All constants in Eq.(\ref{final}) are given in Sec. \ref{S2}. 
In the case of a one component plasma (OCP) consisting of one sort of ions with density $n$ , charges $Ze$ and moving in a neutralizing background all thermodynamical functions may be expressed in terms of the plasma parameter $\Gamma=\beta Z^2 e^2 (4 \pi n/3)^{1/3}$ only. For the excess internal energy defined by
$$
u=\beta \, \frac{\partial \left({\beta F_c/N}\right)}{\partial \beta}
$$
we obtain from Eq.(\ref{final}) 
\beq
u(\Gamma)=p_0 \Gamma^{3/2} + p_1 \Gamma^3 \ln \Gamma + p_2 \Gamma^3 + p_3 \Gamma^{9/2} \ln \Gamma + p_4 \Gamma^{9/2} + p_5 \Gamma^6 \ln^2 \Gamma + p_6 \Gamma^6 \ln \Gamma + p_7 \Gamma^6 \quad,
\eeq
with the constants

\parbox{6.0cm}{
\beq
p_0&=& - \frac{1}{2} \sqrt{3} \quad, \nonumber \\
p_2&=&-\frac{9}{8}\ln 3 - \frac{3}{2} C_E +1 \quad, \nonumber \\ 
p_4&=&0.2350 \quad \nonumber \\
p_6&=&-2.0959  \quad, \nonumber 
\eeq}
\hspace*{2.5cm} \parbox{6.0cm}{
\beq
p_1&=&-\frac{9}{8} \quad, \nonumber \\
p_3&=&-\frac{27}{16}\sqrt{3} \quad , \nonumber \\
p_5&=&- \frac{81}{16} \quad, \nonumber \\
p_7&=&0.0676 \quad.  \nonumber
\eeq}

This expansion completes the result of Ref. \cite{CM59} for the small-$\Gamma$ expansion of the OCP excess internal energy.

In conclusion, in this paper we have studied the low density expansion of the Helmholtz free energy of a classical system of pointlike ions embedded in a neutralizing background and interacting via Coulomb forces. Such a purely Coulomb description is appropriate for a large variety of physical systems. To obtain the virial expansion of a classical Coulomb system the Hubbard-Schofield transformation for the configurational integral in collective variables is used. The original Hamiltonian was maped onto an Ising-like Hamiltonian with harmonic and anharmonic contributions. The coefficients of the new effective Hamiltonian are given in terms of the structure factors of a reference system. In our case with pointlike ions the corresponding reference system is an ideal gas system. 

However, it is possible to generalize the present approach and to consider mixtures of ions with internal structure, i.e., systems with  additional short range interactions. In this case the corresponding reference system is a system with particles interacting solely via short range forces. By introducing short range repulsive interactions $u_{ab}(r)$ (for example hard core repulsions) it is also possible to describe systems of charges with different signs within the present approach. In this case the corresponding cluster expansion is built with the integrable bonds $g_{ab}(r)$, $g^2_{ab}(r)$ and $exp(-\beta u_{ab}(r))\,\, exp(-g_{ab}(r)) - 1 + g_{ab}(r) - (1/2)g^2_{ab}(r)$ 
.

Expanding the anharmonic contributions of the effective Hamiltonian to the configurational integral we have obtained the cluster expansion of a classical Coulomb gas (Eq.(\ref{Zcexpand3})). The exact density expansion of the free energy has been performed up to the order $n^3$ (Eq.(\ref{final})). The virial expansion of other thermodynamic functions can be obtained from (\ref{final}) using thermodynamic identities. Besides its own conceptual interest, the virial expansion might be useful in studying regimes with sufficiently low density and/or sufficiently high temperature. However, using virial expansions for practical calculations one has to take into account that a virial expansion is an asymptotic expansion of the free energy function. The convergence radius of this asymptotic expansion is unknown. Therefore it does not make sense to extent the region of the low density limit to higher densities by calculating higher order terms. We conclude therefore that the obtained virial expansion is applicable for practical calculations only in the low density region (or in the weak coupling regime). 

Finally, the present approach can be generalized to quantum plasmas. Thus one may employ Morita's method of effective potentials to map the quantum plasma system with bare Coulomb interaction to a classical system interacting via effective potentials. The latter split into a Coulomb part and a short range part. Further one expresses the Coulomb interactions through collective variables and maps the original quantum system via the Hubbard-Schofield transformation to a classical reference system with sole short range interactions. Alternatively one may also use the Feynman-Kac formalism. However, explicite calculations to the order $n^3$ are much more complicated as in the classical case.

\section{Acknowledgments}

Valuable discussions with N.~V.~Brilliantov, W.~Ebeling, A.~F\"orster, T.~Kahlbaum and M.~Steinberg are gratefully acknowledged. I would also like to thank I.~M.~Tkachenko and N.~V.~Brilliantov for a carefull reading of the manuscript. 

This work was supported by the Deutsche Forschungsgemeinschaft (Germany).

\begin{appendix}

\section{Calculation of constant ${{b}}_3$ and $b_{4,\alpha}$} \label{AppA}

Consider the leading contribution to the third cluster integral in $k$ space representation
\beq 
\frac{1}{3!}  \,\, 3 \,\,  \begin{minipage} [htbp]{1.3cm} \centerline{\psfig{figure=figures/third1.eps,width=1.1cm,angle=0}}\end{minipage} ~&=&~ {n^3 \, V \over 2} \frac{(-1)^5}{2!\, 2!} \sum_{a,b,c} c_{a}c_{b}c_{c} \,\, l_{ab} l_{ac}^2 l_{bc}^2  \int\!\! \frac{d\vec{k}_{1}d\vec{k}_{2}d\vec{k}_{3}}{\left(2 \pi\right)^9} \,\,\frac{4 \pi}{k_1^2 + \kappa^2} \,\, 
\nonumber \\
&\cdot& \frac{4 \pi}{\left. k_2^2 + \kappa^2 \right.} \,\, \frac{4 \pi}{k_3^2 + \kappa^2} \,\, \frac{4 \pi}{\left(\vec{k}_1-\vec{k}_2\right)^2 + \kappa^2}\,\,\,\frac{4 \pi}{\left(\vec{k}_3-\vec{k}_1\right)^2 + \kappa^2}\,\,\,.
\eeq
Substituting $\vec{x}_i=\vec{k}_i/\kappa$ one expresses the leading term of the third cluster integral through the constant $a_1$ with
\be \label{A2}
b_3~=~\frac{1}{\pi^3} \,\, \int \frac{d x_1 x_1^2}{x_1^2 \,\, +\,\,1} \,\, I^2(\vec{x}_1) \quad ,
\ee
where
\be \label{AI}
I(\vec{y})~=~\int d \vec{x} \frac{1}{x^2+1} \,\, \frac{1}{\left( \vec{x}\,\,+\,\,\vec{y} \right)^2 \,\,+\,\,1} \quad .
\ee
Performing the integration over $\vec{x}$ one obtains
\be \label{AI2}
I(\vec{y})~=~ \frac{2 \pi^2}{y} \,\, \arctan \frac{y}{2} \quad.
\ee
Now in Eq.(\ref{A2}) the integration over $\vec{x}_1$ can be carried out and one obtains \cite{Prudnikov}
\be 
b_3 ~=~ 2 \, \pi^2 \, \left[ \,\, \frac{\pi^2}{12} \,\,-\,\,{1 \over 2} {\ln}^2 {\frac{3}{4}} \,\,-\,\,\mathrm{Li}_2(\frac{1}{4}) \,\, \right] \quad .
\ee
Using the properties of the Euler's dilogarithm one arrives at Eq.(\ref{a_1}).

Consider now the constant $b_{4,\alpha}$ defined in Eq.(\ref{eq:a2alpha}). Using the Fourier representation of $\exp[-x]/x$ we get a similar representation as for $b_{3}$ (Eq.(\ref{A2})),
\be \label{A3}
b_{4,\alpha}~=~\frac{2}{\pi^2} \,\, \int \frac{d x_1 x_1^2}{\left( \, x_1^2 \,\, +\,\,1 \, \right)^2} \,\, I^2(\vec{x}_1) \quad ,
\ee
with the same $I(\vec{y})$ given by Eqs.(\ref{AI}) and (\ref{AI2}). The remaining integration in Eq.(\ref{A3}) leads to the result,
\be
b_{4,\alpha} ~=~ 2 \, \pi^3 \, \left[ \,\, \frac{\pi^2}{12} \,\,-\,\,{1 \over 2} {\ln}^2 {\frac{3}{4}} \,\,-\,\,\mathrm{Li}_2(\frac{1}{4}) \,\,+\frac{4}{3} \ln {\frac{3}{4}} \right] \quad .
\ee
Further transformations lead to the more compact expression Eq.(\ref{a2alpha}).

\section{Calculation of $\delta_1{\cal S}_3$ and $\delta_2{\cal S}_3$} \label{AppB}
After integration over one volume integral and up to the order $n^3$ one represents Eq.(\ref{delta1}) in the following manner,
\be \label{adelta1}
\delta_1{\cal S}_3={V n^3\over 6} \sum_{a,b,c} c_{a}c_{b}c_{c}  \int\!\!d\vec{r}_{1}d\vec{r}_{2}\,\,  \frac{1}{2} \, g_{ab}^2(r_{12}) \,\,\frac{1}{2} \, g_{bc}^2(r_{1}) \,\, \Phi_{ac}(r_{2}) ~+~o\left(n^3\right)
\,\,.
\ee
With the dimensionless variables $\vec{x}_i=\kappa \vec{r}_i$ one has,
\beq
\label{adelta1a}
\delta_1{\cal S}_3=&& \frac{V n^3}{24} \sum_{a,b,c} c_{a}c_{b}c_{c}  l_{ab}^2l_{ac}^2 l_{bc}^2 \int\!\!d{\vec{x}}_{2} \,\left( \,\, \frac{e^{-2 x_2}}{2x_2^2}\,\,-\,\,\frac{l_{ac}\kappa}{3!x_2^3}\, e^{-3x_2} \right.\,\,
\nonumber \\
&+&\left.\,\,\frac{(l_{ac}\kappa)^2}{4!x_2^4}\, e^{-4x_2}\,\,\pm\,\,\dots \,\,\right)\,\,J(x_2) ~+~o\left(n^3\right)\,\,,
\eeq
where
\be
J(x_2)~=~\int d \vec{x_1} \frac{e^{-2x_{12}}}{x_{12}^2}\frac{e^{-2x_1}}{x_1^2}\,\,,
\ee
is representable as
\be
J(\vec{x}_2)~=~\frac{1}{(2\pi)^3}\,\int\, d \vec{k} \,\,\frac{(4\pi)^2}{k^2}\,\arctan^2{\frac{k}{2}}\,\,e^{-i\vec{k}\vec{x_2}}\,\,.
\ee
Within the order $o(n^3)$ we can substitute the contributions with $\exp[-3x_2]/3!x_2^3 \,\,\, \dots$ etc. by their small $x_2$ expansions to get a convergent expression, 

\beq
\delta_1{\cal S}_3=&& \frac{V n^3}{24} \sum_{a,b,c} c_{a}c_{b}c_{c}  l_{ab}^2l_{ac}^2 l_{bc}^2 \int\!\!d{\vec{x}}_{2} \, \,\, \frac{e^{-2 x_2}}{2x_2^2}\,J(x_2)\,\,+\,\,\left( \,\,-\,\,\frac{l_{ac}\kappa}{3!x_2^3}\,\,\, \right.
\nonumber \\
&+& \left. \,\,\frac{(l_{ac}\kappa)^2}{4!x_2^4}\,\,\,\pm\,\,\dots \,\,\right)\,\,J(x_2 \to 0) ~+~o\left(n^3\right)\,\,,
\eeq
We get with $J(x_2 \to 0)~=~\pi^3/x_2$ the following expression
\beq
\delta_1{\cal S}_3=&& \frac{\pi V n^3}{6} \sum_{a,b,c} c_{a}c_{b}c_{c}  l_{ab}^2l_{ac}^2 l_{bc}^2 \lim_{\epsilon \to 0} \lim_{R \to \infty} \int_{\epsilon}^{\kappa R} \!\!d{{x}}_{2} \, \,\, \frac{e^{-2 x_2}}{2}\,J(x_2)\,\,+
\nonumber \\
&+& x_2\,\,\frac{\pi^3}{(l_{ac}\kappa)^2} \,\,\left[ \,\,\exp \left[-\frac{l_{ac}\kappa}{x_2}\right]~-~1~+~\frac{l_{ac}\kappa}{x_2}~-~\frac{1}{2}\frac{(l_{ac}\kappa)^2}{x_2^2} \right]~+~o\left(n^3\right)\,\,.
\eeq
After integration and performing the limiting procedures we obtain the result Eq.(\ref{delta1a}).

Up to the order $n^3$ one represents Eq.(\ref{delta2}) after integration over one volume integral as follows,
\be
\label{adelta2}
\delta_2{\cal S}_3=-\,{V n^3} \sum_{a,b,c} c_{a}c_{b}c_{c} \int\!\!d\vec{r}_{1}d\vec{r}_{2}
 \, g_{ab}(r_{12}) \,\,\frac{1}{2}\, g_{bc}^2(r_{1})  \, \left[ \Phi_{ac}(r_{2})\,\, - \,\, \frac{1}{2} g_{ac}^2(r_{2}) \,\, \right] ~+~o\left(n^3\right)\,\,.
\ee
Introducing dimensionless variables $\vec{x}_i=\kappa \vec{r}_i$ and performing the integration over $\vec{x}_1$ one obtains,
\beq
\label{adelta2a}
\delta_2{\cal S}_3=\,-\, &4& \pi^2 {V n^3} \sum_{a,b,c} c_{a}c_{b}c_{c}  l_{ab}l_{ac}^3 l_{bc}^2 \int_0^{\infty}\!\!d{x}_{2} x_2 \,\left\{ \,\, e^{-x_2}\, \left[\mathrm{Ei}(-x_2)+\ln{3}\,\right] \,\,-\,\,e^{x_2} \, \mathrm{Ei}(-3x_2) \right\}
\nonumber \\
&\cdot&\,\left\{\, \frac{-1}{6} \, \frac{e^{-3x_2}}{x_2^3} \,\,+\,\, \frac{l_{ac} \kappa}{24} \frac{e^{-4x_2}}{x_2^4} \,\,\pm\,\,\dots \,\,\right\}
 \,  ~+~o\left(n^3\right)\,\,.
\eeq 
The higher order terms containing $\exp[-4x_2]/x_2^4 \,\,\, \dots$ can be substituted by their small $x_2$ expansion to cut the singularity at small $x_2$. One gets,
\beq
\label{adelta2b}
\delta_2{\cal S}_3=\,-\,&4& \pi^2 {V n^3} \sum_{a,b,c} c_{a}c_{b}c_{c}  l_{ab}l_{ac}^3 l_{bc}^2 \int_0^{\infty}\!\!d{x}_{2} x_2 \,\left\{ \,\, e^{-x_2}\, \left[\mathrm{Ei}(-x_2)+\ln{3}\,\right] \,\,-\,\,e^{x_2} \, \mathrm{Ei}(-3x_2) \right\}\,\, \frac{-1}{6}\,\frac{e^{-3x_2}}{x_2^3} 
\nonumber \\
&\,&\,\, - \,\, 2 x_2^2 \,\,\, \left( \,\, \frac{l_{ac} \kappa}{4!} \frac{1}{x_2^4} \,\,-\,\,\frac{(l_{ac} \kappa)^2}{5!} \frac{1}{x_2^5}\,\,\pm\,\,\dots \,\,\right) \,\,\left(\, C_E-1+\ln{3x_2}\,\right)
 \,  ~+~o\left(n^3\right)\,\,.
\eeq
Summing up the singular at $x_2=0$ contributions a finite expression is obtained,
\beq
\label{adelta2c}
\delta_2{\cal S}_3&=&\,-\,4 \pi^2 {V n^3} \sum_{a,b,c} c_{a}c_{b}c_{c}  l_{ab}l_{ac}^3 l_{bc}^2 \int_0^{\infty}\!\!d{x}_{2} x_2 \,\left\{ \,\, e^{-x_2}\, \left[\mathrm{Ei}(-x_2)+\ln{3}\,\right] \,\,-\,\,e^{x_2} \, \mathrm{Ei}(-3x_2) \right\}\,\, \frac{-1}{6}\,\frac{e^{-3x_2}}{x_2^3} 
\nonumber \\
\,\,\, &-&\frac{2 x_2^2}{\left(l_{ac}\kappa \right)^3} \,\left(\, C_E-1+\ln{3x_2}\,\right)\,\, \left[ \,\,\exp \left[- \frac{l_{ac} \kappa}{x_2} \right] \,\,-1\,\,+\,\,\frac{l_{ac}\kappa}{x_2}\,\,-\,\,\frac{(l_{ac} \kappa)^2}{2 x_2^2} \,\,+\,\, \frac{(l_{ac} \kappa)^3}{6 x_2^3}  \,\,\right] \,\,
 \,  ~+~o\left(n^3\right)\,\,.
\eeq
The remaining convergent integral can be calculated by introducing an upper and lower bound and then by performing the limiting procedure to $\infty$ and $0$,respectively. After some lengthy calculations one arrives at Eq.(\ref{delta2b}). 

\section{Calculation of $\delta_1{\cal S}_4$} \label{AppC}
The first diagram in Eq.(\ref{fourthdelta}) represents a divergent contribution. Consider therefore the convergent correction to the leading contribution of the first diagram in Eq.(\ref{Fourth})
\be \label{delta1Fourth}
\delta_1{\cal{S}}_{4}~=~ \frac{1}{4!} \,\,\left\{ \,\, 6 \,\,  \begin{minipage} [htbp]{2.3cm} \centerline{\psfig{figure=figures/Fourth1.eps,width=2.1cm,angle=0}}\end{minipage}~-~6 \,\,  \begin{minipage} [htbp]{2.3cm} \centerline{\psfig{figure=figures/fourth1.eps,width=2.1cm,angle=0}}\end{minipage} \right\}\,\,.
\ee
Within the order $o\left(n^3\right)$ one obtains,
\beq
\label{delta1F}
\delta_1{\cal S}_4=&~&{n^4\over 4} \sum_{a,b,c,d} c_{a}c_{b}c_{c}c_{d} \int\!\!d\vec{r}_{a}d\vec{r}_{b}
d\vec{r}_{c}d\vec{r}_d \, 2 \, \left[ \Phi_{ab}(r_{ab})\,\, - \,\, \frac{1}{2} g_{ab}^2(r_{ab}) \,\, \right] 
\nonumber\\
&\,& g_{ac}(r_{ac}) \,  \,g_{bd}(r_{bd}) \,  \, \frac{1}{2} g_{cd}^2(r_{cd})~+~o\left(n^3\right) \,.
\eeq
or
\beq
\label{adelta1F}
\delta_1{\cal S}_4=~&-&~ \frac{V n^4}{4 \kappa^4} \beta^7 \sum_{a,b,c,d} c_{a}c_{b}c_{c}c_{d} e_a^4e_b^4e_c^3e_d^3 \int\!\!d{\vec{x}} \,\, \frac{e^{-3 x}}{6x^3}\,J_4(x)\,\,-\,\,\left( \,\,\frac{l_{ab}\kappa}{4!x^4}\,\,\, \right.
\nonumber \\
&+& \left. \,\,\frac{(l_{ac}\kappa)^2}{5!x^5}\,\,\,\pm\,\,\dots \,\,\right)\,\,J_4(x \to 0) ~+~o\left(n^3\right)\,\,,
\eeq
where
\be
J_4(x_1)~=~\int d \vec{x_2} d \vec{x_3} \frac{e^{-2x_{23}}}{x_{23}^2}\frac{e^{-x_2}}{x_2}\frac{e^{-x_{13}}}{x_{13}}\,\,,
\ee
is representable as
\be
J_4(\vec{x})~=~\frac{1}{(2\pi)^3}\,\int\, d \vec{k} \,\,\frac{(4\pi)^2}{(k^2+1)^2}\,\frac{4\pi}{k} \arctan{\frac{k}{2}}\,\,e^{-i\vec{k}\vec{x}}\,\,.
\ee
Taking into account that $J_4(0)=8 \pi^2/3$ we get the convergent expression
\beq
\delta_1{\cal S}_4=~&-&~ \frac{\pi V n^4}{\kappa^2} \beta^7 \sum_{a,b,c,d} c_{a}c_{b}c_{c}c_{d} e_a^4e_b^4e_c^3e_d^3  \lim_{\epsilon \to 0} \lim_{R \to \infty} \int_{\epsilon}^{\kappa R} \!\!d{{x}} \, \,\, \frac{e^{-3 x}}{6x}\,J_4(x)\,\,-
\nonumber \\
&-& x^2\,\,\frac{8 \pi^2}{3 (l_{ab}\kappa)^3} \,\,\left\{ \,\,\exp \left[-\frac{l_{ab}\kappa}{x}\right]~-~1~+~\frac{l_{ab}\kappa}{x}~-~\frac{1}{2}\frac{(l_{ab}\kappa)^2}{x^2}~+~\frac{1}{6}\frac{(l_{ab}\kappa)^3}{x^3} \right\}~+~o\left(n^3\right)\,\,.
\eeq
After integration and performing the limiting procedures one gets the logarithmic contribution of Eq.(\ref{deltafourth}) and the constant $c_{4\alpha}$ (Eq.(\ref{eq:c_4alpha})).
\end{appendix}

\end{document}